\documentclass[12pt]{article}
\RequirePackage{amsthm,amsmath,amsbsy,amsfonts}
\usepackage{bm,graphicx,psfrag,epsf}
\usepackage{enumerate}
\usepackage{natbib}
\RequirePackage[colorlinks,citecolor=blue,urlcolor=blue]{hyperref}
\RequirePackage{hypernat}
\usepackage{paralist, booktabs, multirow}
\usepackage{algorithm,algorithmic}
\usepackage{url}

\newcommand{\blind}{0}

\usepackage[mathscr]{eucal}

\numberwithin{equation}{section}
\theoremstyle{plain}
\newtheorem{thm}{Theorem}[section]

\newtheorem{remark}{Remark}

\newcommand{\V}[1]{{\bm{\mathbf{\MakeLowercase{#1}}}}} % vector
\newcommand{\Vhat}[1]{{\bm{\hat{\mathbf{\MakeLowercase{#1}}}}}} % vector
\newcommand{\VE}[2]{\MakeLowercase{#1}_{#2}} % vector element
\newcommand{\Vtilde}[1]{{\bm{\tilde{\mathbf{\MakeLowercase{#1}}}}}} % vector
\newcommand{\Vn}[2]{\V{#1}^{(#2)}} % n-th vector

\newcommand{\Real}{{\mathbb R}}
\newcommand{\Tra}{^{{\sf T}}} % transpose
\newcommand{\Inv}{^{-1}} % inverse

\newcommand{\M}[1]{{\bm{\mathbf{\MakeUppercase{#1}}}}} % matrix
\newcommand{\Mtilde}[1]{{\bm{\tilde{\mathbf{\MakeUppercase{#1}}}}}} % vector
\newcommand{\ME}[2]{\MakeLowercase{#1}_{#2}} % matrix element

\begin{document}

\def\spacingset#1{\renewcommand{\baselinestretch}%
{#1}\small\normalsize} \spacingset{1}

%%%%%%%%%%%%%%%%%%%%%%%%%%%%%%%%%%%%%%%%%%%%%%%%%%%%%%%%%%%%%%%%%%%%%%%%%%%%%%

\if0\blind
{
  \title{\bf Robust Parametric Classification and Variable Selection
by a Minimum Distance Criterion}
  \author{Eric C. Chi\thanks{
    Eric C. Chi (E-mail: ecchi@ucla.edu) is Postdoctoral Scholar, Department of Human Genetics, University of California, Los Angeles CA 90095-7088.} \,and
%    and
    David W. Scott\thanks{
    David W. Scott (E-mail: scottdw@rice.edu) is Professor, Department of Statistics, Rice University, Houston, TX 77005.}\\}
  \maketitle
} \fi

\if1\blind
{
  \bigskip
  \bigskip
  \bigskip
  \begin{center}
    {\LARGE\bf Title}
\end{center}
  \medskip
} \fi

\bigskip
\begin{abstract}
We investigate a robust penalized logistic regression algorithm based on a minimum distance criterion. Influential outliers are often associated with the explosion of parameter vector estimates, but in the context of standard logistic regression, the bias due to outliers always causes the parameter vector to implode, that is shrink towards the zero vector. Thus, using LASSO-like penalties to perform variable selection in the presence of outliers can result in missed detections of relevant covariates.  We show that by choosing a minimum distance criterion together with an Elastic Net penalty, we can simultaneously find a parsimonious model and avoid estimation implosion even in the presence of many outliers in the important small $n$ large $p$ situation.  Minimizing the penalized minimum distance criterion is a challenging problem due to its nonconvexity. To meet the challenge, we develop a simple and efficient MM algorithm that can be adapted gracefully
to the small $n$ large $p$ context. Performance of our algorithm is evaluated on simulated and real data sets.
This article has supplementary materials online.
\end{abstract}

\noindent%
{\it Keywords:}  {Logistic regression, Robust estimation, Implosion breakdown, LASSO, Elastic Net, Majorization-Minimization}

\spacingset{1.45}
\section{Introduction}
\label{sec:intro}

Regression, classification and variable selection problems in high dimensional data are becoming routine in fields ranging from finance to genomics. In the latter case, technologies such as expression arrays have made it possible to comprehensively query a patient's transcriptional activity at a cellular level. Patterns in these profiles can help refine subtypes of a disease according to sensitivity to treatment options or identify previously unknown genetic components of a disease's pathogenesis.

The immediate statistical challenge is finding those patterns when the number of predictors far exceeds the number of samples. To that end the Least Absolute Shrinkage and Selection Operator (LASSO) has been quite successful at addressing ``the small $n$, big $p$ problem" (\citealp*{Tibshirani1996}; \citealp{Chen1998}). Indeed, $\ell_1$-penalized maximum likelihood model fitting has inspired many related approaches that simultaneously do model fitting and variable selection. These approaches have been extended from linear regression to generalized linear models. In particular, linear models minimizing the logistic deviance loss with an Elastic Net penalty \citep{Zou2005} have been well studied
(\citealp{Genkin2007}; \citealp{Liu2007}; \citealp{Wu2009}; \citealp{Friedman2010})

Nonetheless while $\ell_1$-penalized maximum likelihood methods have proved their worth at recovering parsimonious models, less attention has been given to extending these methods to handle outliers in high dimensional data. For example in biological data, tissue samples may be mislabeled or be contaminated. The majority of prior work centers on linear regression (\citealp{Rosset2007}; \citealp{Wang2007}; \citealp{Li2011}; \citealp{Alfons2012}), although there are a few exceptions. 
\citet{Rosset2007} and \citet*{Wang2008} discuss using a Huberized hinge loss for regularized classification, and \citet{Geer2008} studies LASSO penalization of generalized linear models.
Nonetheless, with the exception of the $\ell_1$-penalized least trimmed squares regression procedure of \citet{Alfons2012} and the Huberized hinge loss, these approaches can provide robustness only to outliers in the response variable, not to outliers in the covariates. Moreover, neither paper on the Huberized hinge loss is primarily concerned with robustness. \cite{Rosset2007} present impressive general conditions that ensure piecewise linear regularization paths. The Huberized hinge loss is introduced as an illustration and applied on a small example that highlights its prediction accuracy in the presence of a single gross outlier. Despite being introduced as a loss for a robust procedure in \citet{Rosset2007}, the primary motivation for using the Huberized hinge loss in \citet{Wang2008} is the fast algorithm introduced in \citet{Rosset2007} for computing the entire regularization path, not its robustness properties. We will see later that this loss can struggle under a heavy dose of outliers.

Robustness against outlying covariate values warrants further investigation. It is not surprising that outliers may bias estimation.  What is less well appreciated is that outliers can strongly influence variable selection.  In this paper we identify some circumstances that motivate robust variants of penalized estimation and develop a minimum distance estimator for logistic regression. To address the $n \ll p$ scenario when predictors are correlated we add the Elastic Net penalty. We evaluate the performance of our approach through simulated and real data.

Robust methods of logistic regression are not new in the classic $n > p$ case.  A broad class of solutions consists of downweighting the contribution of outlying points to the estimating equations. Downweighting can be based on extreme values in covariate space (\citealp{kunsch1989}; \citealp{carrol1993}) or on extreme predicted probabilities (\citealp{copas1988}; \citealp{carrol1993}; \citealp{bianco1996}).

An alternative approach is to use minimum distance estimation \citep{Donoho1988}. The minimum distance estimator used in this paper can also be seen as a method that downweights the contributions of outliers \citep{Chi2011}. The work in \citet{bondell2005} is similar to ours in that he considered fitting parameters by minimizing a weighted  Cram\'{e}r-von Mises distance. The difference between the approach proposed here and prior work is the application of regularization to handle high dimensional data and perform variable selection in the presence of outliers. Moreover, the robust loss function we propose has a particularly simple form which, when combined with the Elastic Net penalty, can be solved very efficiently for large problems by minimizing a series of penalized least squares problems with coordinate descent.

The rest of this paper is organized as follows.
In Section~\ref{sec:mle} we review maximum likelihood estimation (MLE) of the logistic regression model and demonstrate the potentially deleterious effects of outliers on variable selection with the $\ell_1$-penalized MLE. We introduce our robust loss function in Section~\ref{sec:L2E}. In Section~\ref{sec:algorithm} we describe algorithms for fitting our robust logistic regression model. In Sections~\ref{sec:simulations} and \ref{sec:real_data} we present results on real and simulated data. Section~\ref{sec:discussion} concludes with a summary of our work and also future directions.

\section{Standard logistic regression and implosion breakdown}
\label{sec:mle}
Throughout this paper we adopt the following conventions. We assume that the columns of the design matrix $\M{X}$ are centered. We overload notation so that if $f$ is a function of a scalar, then $f$ evaluated at vector or matrix should be interpreted as being evaluated element-wise. For a linear model $\beta_0\V{1} + \M{X}\V{\beta}$ we will often employ the compact notations $\Mtilde{X} = (\V{1}, \M{X}) \in \Real^{n \times (p+1)}$ and 
$\V{\theta} = (\beta_0, \V{\beta}\Tra)\Tra \in \Real^{p+1}$.

\begin{figure}
	\centering
		\includegraphics[scale=0.705]{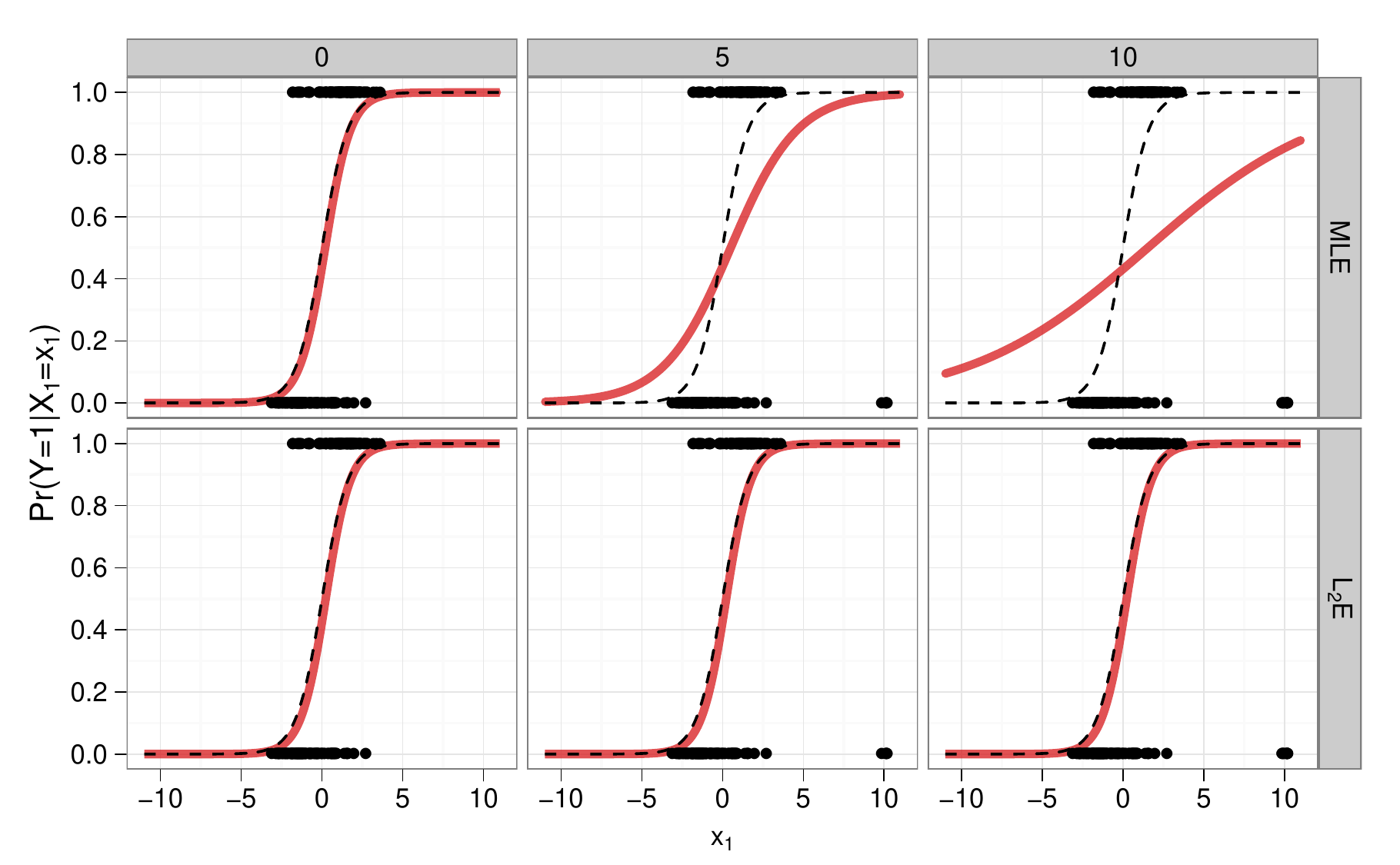}
	\caption{Univariate regression onto $X_1$. The dashed line denotes the
    logistic model that generated the data; the heavy solid line denotes the estimated response. The number of outliers (0, 5, 10) increases from left to right. The first row shows MLE results; the second shows L$_2$E results.
    \label{fig:univariate}}	
\end{figure}

In binary regression, we seek to predict or explain an observed response $\V{y} \in \{0,1\}^n$ using predictors $\M{X}\in\Real^{n\times p}$,
where $n\ll p$ may be expected.  In typical expression microarray data we encounter $n\approx 100$ and
$p\approx 10^4$, while with single nucleotide polymorphism (SNP) array data both $n$ and $p$ may be larger by a factor of 10. Let the conditional probabilities be given by $P(Y_i = 1 | X_i = \V{x}_i) = F(\Vtilde{X}_i\Tra\V{\theta})$ where $F(u)=1/(1+\exp(-u))$. Then under this assumption, in standard logistic regression \citep{McCullagh1989} we minimize the negative log-likelihood of a linear summary of the predictors,
\begin{equation}
	\label{eq:lr}
	\V{y}\Tra\Mtilde{X}\V{\theta} - \V{1}\Tra\log (\V{1} + \exp(\Mtilde{X}\V{\theta})).
\end{equation}
A simple univariate example illustrates the bias that outliers can introduce into this estimation procedure. In the top panel of Figure~\ref{fig:univariate} we see that the addition of 5 and 10 outliers among the
controls shrinks $\Vhat{\beta}$ towards zero.  In fact, \citet{Croux2002} showed that with $p$ covariates only $2p$ such outliers are required to make $\lVert \hat {\V{\beta}} \rVert_2<\epsilon$ for any desired $\epsilon$. 
Our robust estimator, which we introduce in the next section, produces virtually the same curves shown in the bottom panel of Figure \ref{fig:univariate}.

\begin{figure}
	\centering
		\includegraphics[scale=0.705]{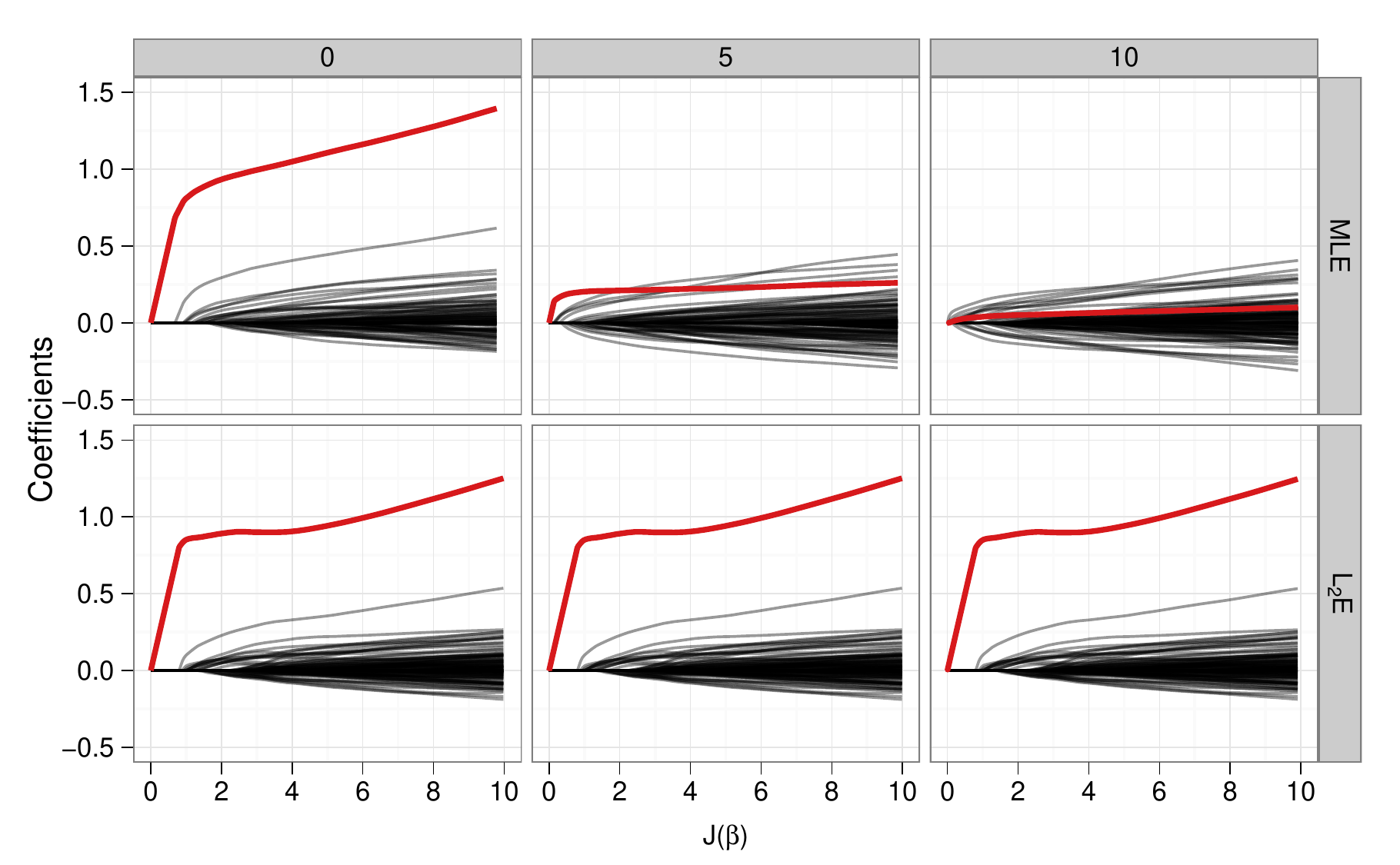}
	\caption{Regularization paths. The heavy line denotes the path for the relevant regression coefficient $\beta_1$; $J(\V{\beta})$ is the 1-norm of $\V{\beta}$. The number of outliers (0, 5, 10) increases from left to right; 999 irrelevant covariates have been added.  The first row shows MLE results; the second shows L$_2$E results.\label{fig:multivariate}}
\end{figure}

This ``implosion" breakdown phenomenon has implications for LASSO based variable selection. Consider what happens when we add 999 noise covariates  which are independent of the class labels to the scenario depicted in Figure~\ref{fig:multivariate} and perform $\ell_1$-penalized logistic regression.  The top panel of Figure~\ref{fig:multivariate} shows the corresponding regularization paths or the values of the fitted regression coefficients as a function of the penalization parameter. As outliers are added the regularization path for the relevant covariate $X_1$ quickly falls into the noise.

The LASSO performs continuous variable selection by shrinking to zero regression coefficients of covariates with very low correlation with the responses. If outliers are present in relevant covariates, then the combination of implosion breakdown and soft-thresholding by the LASSO can lead to missed detection of relevant covariates. In contrast we see 
in the bottom panel of Figure~\ref{fig:multivariate} that the corresponding regularization paths obtained using our robust estimator are insensitive to outliers and so relevant covariates still have the chance of being selected. This simple example highlights the potential importance of penalized robust estimation procedures. In the next section we describe our robust estimator.

\section{The Minimum Distance Estimator}
\label{sec:L2E}
Let $P_{\V{\theta}}$ be a probability mass function (PMF), specified by a parameter $\V{\theta} \in \Theta \subset \Real^p$, believed to be generating data $Y_1, \ldots, Y_n$ that take on values in the discrete set $\chi$. Let $P$ be the unknown true PMF generating the data. If we actually knew the
true distribution, an intuitively good solution is the one that is ``closest" to the true distribution. Consequently, as an alternative to using the negative log-likelihood, we consider the L$_2$ distance between $P_{\V{\theta}}$ and $P$. Thus, we pose the following variational optimization problem; we seek $\Vhat{\theta} \in \Theta$ that minimizes
\begin{equation}
\label{eq:ISE}
\sum_{y \in \chi} \left [P_\V{\theta}(y) - P(y) \right ]^2.
\end{equation}
Although finding such a $\V{\theta}$ is impossible since $P$ is unknown, it is possible to find a $\V{\theta}$ that minimizes an unbiased estimate of this distance.
Expanding the sum in (\ref{eq:ISE}) gives  us
\begin{equation*}
\label{eq:discreteL2E}
\sum_{y\in\chi} P_{\V{\theta}}(y)^2   - 2\sum_{y\in\chi} P_{\V{\theta}}(y)P(y) +  \sum_{y\in\chi} P(y)^2.
\end{equation*}
The second summation is an expectation $E[P_{\V{\theta}}(Y)]$ where $Y$ is a random variable drawn from $P$. This summation can be estimated from the data by the sample mean. The third summation does not depend on $\V{\theta}$. With these observations in mind, we  use the following fully data-based loss function
\begin{equation}
\label{eq:l2e}
L(\V{\theta}) = \sum_{y\in\chi} P_{\V{\theta}}(y)^2 - \frac{2}{n}\sum_{i=1}^n P_{\V{\theta}}(y_i)
\end{equation}
and seek a $\Vhat{\theta}$ such that $L(\Vhat{\theta}) = \min_{\V{\theta}\in\Theta} L(\V{\theta})$. The estimate $\Vhat{\theta}$ is called an L$_2$ estimate or L$_2$E in \citet{Scott2001}.

The above minimization problem is a familiar one associated with bandwidth selection for histograms and more generally for kernel density estimators \citep{Scott2008}. Applying a commonly used criterion in nonparametric density estimation to parametric estimation has the interesting consequence of trading off efficiency with robustness in the estimation procedure.  In fact, previously \citet{Basu1998} introduced a family of divergences which includes the L$_2$E as a special case and the MLE as a limiting case. The members of this family of divergences are indexed by a parameter that explicitly trades off efficiency for robustness. The MLE is the most efficient but least robust member in this family of estimation procedures. The L$_2$E represents a reasonable tradeoff between efficiency and robustness. \citet{Scott2001,Scott2004} demonstrated that the L$_2$E has two benefits, the aforementioned robustness properties and computational tractability. The tradeoff in asymptotic efficiency is similar to that seen in comparing the mean and median as a location estimator. Indeed, while other members in this family may possess a better tradeoff, the L$_2$E has the advantage of admitting a simple and fast computational solution as we will show in Section~\ref{sec:algorithm}.

We now show that the L$_2$E method applied to logistic regression amounts to solving a non-linear least squares problem. We seek to minimize a surrogate measure of the L$_2$ distance between the logistic conditional probability and the conditional probability generating the data. If the $\V{x}_i$ are unique, then $y_i \sim \textsc{B}(1,p_i)$ where $p_i = F(\Vtilde{x}_i\Tra\V{\theta})$. The L$_2$E loss for this one sample is $p_i^2+(1-p_i )^2  -  2 [  y_i p_i  + (1-y_i)(1-p_i ) ]$. Extending to the entire sample, a sensible approach is to minimize the average L$_2$ distance, namely
\begin{equation}
\label{eq:logisticL2E_loss}
\begin{split}
	\frac{1}{n}\sum_{i=1}^n \left [p_i^2+(1-p_i )^2  -  2 [  y_i p_i  + (1-y_i)(1-p_i ) ] \right].
\end{split}
\end{equation}
Up to an additive constant that does not depend on $\V{\theta}$, the criterion in (\ref{eq:logisticL2E_loss}) can be compactly written as
\begin{equation*}
\label{eq:logisticL2E_loss_Compact}
	L(\V{y}, \Mtilde{X}\V{\theta}) = \frac{1}{n} \lVert \V{y} - F(\Mtilde{X}\V{\theta}) \rVert_2^2,
\end{equation*}
after dividing by two. Remarkably, minimizing this unassuming loss function produces robust logistic regression coefficients. A closer inspection of the estimating equations gives some intuition for the
logistic L$_2$E's robustness. A stationary point $\V{\theta}^*$ of the L$_2$E loss satisfies
\begin{equation*}
0 = \sum_{i=1}^n \gamma_i^* \V{x}_i [y_i - F(\Vtilde{X}_i\Tra\V{\theta}^*)]
\end{equation*}
where $\gamma_i^* =  F(\Vtilde{X}_i\Tra\V{\theta}^*)[1-F(\Vtilde{X}_i\Tra\V{\theta}^*)]$. Thus, at a stationary point $\V{\theta}^*$, the discrepancies between observed and fitted values, namely $y_i - F(\Vtilde{X}_i\Tra\V{\theta}^*)$, are small for samples with predicted values that are far from the extreme values of one and zero, namely samples for which $\gamma_i^*$ are not close to zero. The $i$th discrepancy is free to be large for samples with predicted values close to zero or one, namely samples for which $\gamma_i^*$ are close to zero. Very large and small predicted values tend to occur at extreme values of the covariates given the sigmoid shape of $F$. Thus, observations that are extreme in the covariate space contribute very little to the estimating equations at $\V{\theta}^*$. Moreover, we see that the robustness does not rely on $F$ being the logistic link; rather we just require that $F$ be sigmoid. Finally, we note that the estimating equations also show us that the L$_2$E is affine equivariant, namely linear transformations of the covariates change the estimated regression coefficients accordingly, and therefore linear transformations of the covariates do not change the fitted responses. For more in depth discussion on the theory behind minimum distance estimators like the L$_2$E, we refer readers to the works of \citet{Basu1998} and \cite{Donoho1988}.

Before moving on to discuss our algorithm, we remark that the L$_2$ distance has been used before for classification problems. \citet{Kim2008, Kim2010} used the L$_2$ distance to perform classification using kernel density estimates. Their application of the L$_2$ distance, however, is more in line with its customary use in nonparametric density estimation whereas we use it to robustly fit a parametric model.

\section{Estimation with convex quadratic majorizations}
\label{sec:algorithm}
We now derive an algorithm for finding the logistic L$_2$E solution by minimizing a series of convex quadratic losses. 
We minimize the L$_2$E loss with a Majorization-Minimization (MM) algorithm (\citealp*{Lange2000}; \citealp*{Hunter2004}) because it is numerically stable and easy to implement. Most importantly, our MM algorithm is also easily adapted to handle LASSO-like penalties.

%\subsection{Majorization-Minimization}
The strategy behind MM algorithms is to minimize a surrogate function, the majorization, instead of the original objective function. The surrogate is chosen with two goals in mind. First, an argument that decreases
the surrogate should decrease the objective function. Second, the surrogate should be easier to minimize than the objective function. Formally stated, a real-valued function $h$ majorizes a real-valued function $g$ at $\mathbf{v}$ if $h(\mathbf{u}) \geq g(\mathbf{u})$ for all $\mathbf{u}$ and $h(\mathbf{v}) = g(\mathbf{v})$.
Given a procedure for constructing a majorization, we can define the MM algorithm to find a minimizer of a function $g$  as follows. Let $\Vn{v}{k}$ denote the $k$th iterate: 
\begin{inparaenum}[(1)]
  \item find a majorization $h(\V{v}; \Vn{v}{k})$ of $g$ at $\Vn{v}{k}$;
  \item set $\Vn{v}{k+1} = \arg \min_{\V{v}} h(\V{v}; \Vn{v}{k})$; and
  \item repeat until convergence.
\end{inparaenum}
This algorithm always takes non-increasing steps with respect to $g$. By using the MM algorithm, we can convert a hard optimization problem into a series of simpler ones, each of which is easier to minimize than the original.

To estimate $\Vhat{\theta}$ such that $L(\V{y}, \Mtilde{X}\Vhat{\theta}) = \min_{\V{\theta}} L(\V{y}, \Mtilde{X}\V{\theta})$ we rely on the following convex quadratic majorization.
\begin{thm}
	\label{thm:MM}
	The following function majorizes $L(\V{y}, \Mtilde{X}\V{\theta})$ at $\Vtilde{\theta}$:
	\begin{equation}
	\label{eq:logisticL2Emajorization}
	\begin{split}
		L(\V{\theta};\Vtilde{\theta}) = L(\V{y}, \Mtilde{X}\Vtilde{\theta}) + \frac{2}{n} \V{z}_\Vtilde{\theta} \Tra\Mtilde{X}(\V{\theta}-\Vtilde{\theta})
+ \frac{\eta}{n} \lVert \Mtilde{X}(\V{\theta} - \Vtilde{\theta}) \rVert_2^2, \\
	\end{split}
	\end{equation}
	where $\V{z}_\Vtilde{\theta} = 2\M{G}[F(\Mtilde{X}\Vtilde{\theta}) - \V{y}]$, $\M{G}$ is diagonal with
	$\ME{G}{ii} = F(\Vtilde{X}_i\Tra\Vtilde{\theta})[1-F(\Vtilde{X}_i\Tra\Vtilde{\theta})]$,
	and $\eta > 0$ is sufficiently large.
\end{thm}
Using the majorization (\ref{eq:logisticL2Emajorization}) in an MM algorithm results in iterative least squares. A proof of Theorem~\ref{thm:MM} is given in the Supplementary Materials. We are able to find a simple convex quadratic majorization since the logistic L$_2$E loss has bounded curvature. A sharp lower bound on $\eta$ is given by the maximum curvature of the logistic L$_2$E loss over all parameter values. The bound is derived in the Supplementary Materials. The practical implication is that  the parameter $\eta^{-1}$ controls the step size of our iterative solver. Consequently, in practice we set $\eta$ to its lower bound to take the largest steps possible to speed up convergence.

We can express the majorization $L(\V{\theta},\Vtilde{\theta})$ in (\ref{eq:logisticL2Emajorization}) as
\begin{equation*}
	L(\V{\theta},\Vtilde{\theta}) = \eta(\tilde{\beta}_0 - \beta_0 - \frac{1}{\eta} \overline{z}_{\Vtilde{\theta}} )^2
	+ \frac{\eta}{n}\lVert \zeta(\Vtilde{\theta}) - \M{X}\V{\beta}\rVert_2^2
	 + K(\Vtilde{\theta}),
\end{equation*}
where $\overline{z}_\Vtilde{\theta} = n\Inv\V{1}\Tra\V{z}_\Vtilde{\theta},$
	$\zeta(\Vtilde{\theta}) = \M{X}\Vtilde{\beta} - \eta^{-1}(z_{\Vtilde{\theta}} - \overline{z}_\Vtilde{\theta}\V{1})$,
and $K(\Vtilde{\theta})$ is a constant that does not depend on $\V{\theta}$. 
When $\M{X}$ is full rank, as is often the case when $n > p$, then the solution to the normal equations is unique and the parameter updates are given by 
\begin{equation}
\label{eq:IRLS}
\begin{split}
	\beta_0^{(m+1)} &= \beta_0^{(m)} - \eta^{-1} \overline{z}_{\V{\theta}^{(m)}}, \\
	\Vn{\beta}{m+1} &= \Vn{\beta}{m} - \frac{1}{\eta} \left ( \M{X}\Tra\M{X}\right )\Inv\M{X}\Tra\V{z}_{\V{\theta}^{(m)}}. \\
\end{split}
\end{equation}
The descent direction has a simple update since the Hessian approximation is computed only once for all iterations.

The majorization given in Theorem~\ref{thm:MM} can be adapted for regularization. It follows immediately that $(1/2)L(\V{\theta};\Vtilde{\theta}) + \lambda J(\V{\beta})$ majorizes $(1/2)L(\V{y}, \Mtilde{X}\V{\theta}) + \lambda J(\V{\beta})$ for a penalty function $J: \mathbb{R}^{p} \rightarrow \mathbb{R}_+$ and positive regularization parameter $\lambda$. Note that the intercept parameter is not penalized. Regularization is useful for stabilizing estimation procedures. For example, if $\M{X}$ is not full rank or has a large condition number, a ridge penalty can salvage the situation. We then seek the minimizer to the following problem
\begin{equation*}
	\min_{\V{\theta} \in \Real^{p+1}} \frac{1}{2n}\lVert \V{y} - F(\Mtilde{X}\V{\theta}) \rVert_2^2 + \lambda \frac{1}{2}\lVert \V{\beta} \rVert_2^2,
\end{equation*}
which we can solve by minimizing the majorization $L(\V{\theta}, \Vtilde{\theta}) + \lambda \lVert \V{\beta} \rVert_2^2$. Since the intercept is not penalized, the intercept updates are the same as in (\ref{eq:IRLS}). The update for $\V{\beta}$ becomes
\begin{equation}
\label{eq:IRLS_ridge}
	\Vn{\beta}{m+1} = \Vn{\beta}{m} - \frac{1}{\eta}(\M{X}\Tra\M{X} + \lambda\M{I})\Inv \M{X}\Tra\V{z}_{\V{\theta}^{(m)}}. \\
\end{equation}

Under suitable regularity conditions, the MM algorithm for solving the ridge penalized logistic L$_2$E problem is guaranteed to converge to a stationary point of $L(\V{y}, \Mtilde{X}\V{\theta}) + \lambda\lVert \V{\beta} \Vert_2^2$. This follows from global convergence properties of MM algorithms that involve continuously differentiable objective and majorization functions \citep{Lange2010}. On the other hand, the MM algorithm for the unregularized version of the problem is not guaranteed to converge based on the sufficient conditions given in \cite{Lange2010} because the objective function is not coercive (i.e.\@, not all its level sets are compact) and the quadratic majorization is not strictly convex in $\V{\theta}$ unless $\M{X}$ is full rank. Adding the ridge penalty remedies both situations, and sufficient conditions for global convergence are met.

Another reason to consider regularization is to perform continuous variable selection via a LASSO-like penalty.  In particular, consider the penalized majorizer for the L$_2$E loss regularized by the Elastic Net penalty, $J(\V{\beta}) =  \lambda \left (\alpha \lVert \V{\beta} \rVert_1+ (1-\alpha)/2\lVert \V{\beta} \rVert_2^2 \right )$ where $\alpha \in [0,1]$ is a mixing parameter between the ridge and LASSO penalty.
Since our work is motivated by genomic data which are known to have correlated covariates, we will focus on the Elastic Net penalty because it produces sparse models but includes and excludes groups of correlated variables \citep{Zou2005}. The LASSO, in contrast, tends to select one covariate among a group correlated covariates and exclude the rest. If groupings among the covariates are known in advance, a group LASSO penalty could be used \citep{Yuan2006}. The Elastic Net penalty is useful in that it performs group selection without prespecification of the groups. Thus, we are interested in generating MM iterates $\Vn{\theta}{m} = \left (\beta_0^{(m)}, \Vn{\beta}{m} \right)$ where
\begin{equation}
\label{eq:ENLS}
\begin{split}
	\beta_0^{(m+1)} &= \beta_0^{(m)} - \eta^{-1} \overline{z}_{\V{\theta}^{(m)}}\\
	\Vn{\beta}{m+1} &= \underset{\V{\beta} \in \Real^p}{\arg\min}\; \frac{\eta}{2n}\lVert \zeta(\Vn{\theta}{m}) - \M{X}\V{\beta}\rVert_2^2 + \lambda \left (\alpha \lVert \V{\beta} \rVert_1+ \frac{(1-\alpha)}{2}\lVert \V{\beta} \rVert_2^2 \right ). \\
\end{split}
\end{equation}

Before discussing how to practically solve the surrogate minimization problem, note that regardless of how (\ref{eq:ENLS}) is solved, we have the following guarantee on the convergence of the MM iterates.
\begin{thm}
\label{thm:convergence_n_g_p}
Under suitable regularity conditions, for any starting point $\Vn{\theta}{0}$, the sequence of iterates $\Vn{\theta}{1}, \Vn{\theta}{2}, \ldots$ generated by (\ref{eq:ENLS}) converges to a stationary point of
\begin{equation*}
	\frac{1}{2n}\lVert \V{y} - F(\Mtilde{X}\V{\theta})\rVert_2^2 + \lambda \left (\alpha \lVert \V{\beta} \rVert_1+ \frac{(1-\alpha)}{2}\lVert \V{\beta} \rVert_2^2 \right ),
\end{equation*}
where $\lambda > 0$ and $\alpha \in [0, 1)$.
\end{thm}
A proof is given in the Supplementary Materials and relies on an extension of the global convergence properties of MM algorithms for locally Lipschitz continuous objective and majorization functions \citep{Schifano2010}. 
Note that Theorem~\ref{thm:convergence_n_g_p} restricts $\alpha < 1$, i.e.\@, algorithmic convergence of the LASSO regularized logistic L$_2$E is not guaranteed. This  condition is imposed to ensure that the majorization is strictly convex in $\V{\beta}$. In our experience, the LASSO regularized logistic L$_2$E does not have algorithmic convergence issues in practice.
As a final remark on algorithmic convergence, note that since the ridge penalty is a special case of the Elastic Net, Theorem~\ref{thm:convergence_n_g_p} implies that ridge penalized logistic L$_2$E (\ref{eq:IRLS_ridge}) will also converge.

To solve (\ref{eq:ENLS}) we turn to coordinate descent which has been shown to efficiently solve penalized regression problems when selecting relatively few groups of correlated predictors \citep*{Friedman2007, Wu2008}.
Coordinate descent is a special case of block relaxation optimization where, in a round-robin fashion, we optimize the objective function with respect to each coordinate at a time while holding all other coordinates fixed. 

The $j$th coordinate update during the $k$th round of coordinate descent of the $m$th MM iteration, denoted
$\VE{\beta}{j}^{(m,k)}$, has a simple form \citep{Donoho1995} and is given by the subgradient equations to be
\begin{equation*}
	\VE{\beta}{j}^{(m, k)} = \frac{S\left(\frac{\eta}{n}\V{X}_{(j)}\Tra\V{r}^{(m, k, j)} , \lambda\alpha\right)}
{\frac{\eta}{n}\lVert \V{x}_{(j)} \rVert_2^2 + \lambda(1-\alpha)},
\end{equation*}
where $\V{x}_{(j)}$ denotes the $j$th column of $\M{X}$ and $\V{r}^{(m, k, j)}$ is a vector of partial residuals with $i$th entry
\begin{equation*}
	\VE{r}{i}^{(m, k, j)} = \VE{\zeta}{i} ( \Vn{\theta}{m})- \left(\sum_{j'=1}^{j-1}\ME{X}{ij'}\VE{\beta}{j'}^{(m, k)} + \sum_{j'=j+1}^p\ME{X}{ij'}\VE{\beta}{j'}^{(m, k-1)}\right),
\end{equation*}
and $S$ is the soft-threshold function: $S(a,\lambda) = \operatorname{sign}(a)\max(|a| - \lambda,0).$ Additional details on how coordinate descent is nested within the MM steps and how convergence is evaluated can be found in the Supplementary Materials.

\section{Simulations}
\label{sec:simulations}
In this section we report on three simulations comparing the MLE and L$_2$E results. The first two simulations examine the accuracy of estimation. We then follow with a simulation experiment designed to examine the variable selection properties.  For the first two simulations we generated $1000$ data sets, with $200$ binary outcomes each associated with $4$ covariates, from the logistic model specified by the likelihood in (\ref{eq:lr}) with parameters $\beta_0=0$ and $\V{\beta} = (1, 0.5,1,2)\Tra$. The covariates $\V{x}_i$ were drawn from one of two populations.
For $i = 1, \ldots, 100$, the $\V{x}_i$ are i.i.d samples from $N(\V{\mu}, 0.16\, \M{I}_p)$ and for $i = 101, \ldots 200$,
they are i.i.d samples from $N(-\V{\mu}, 0.16 \, \M{I}_p)$,
where $p = 4$ and $\V{\mu} = (0.25, 0.25, 0.25, 0.25)\Tra$. The responses were generated independently as $y_i\sim \textsc{B}(1,F(\V{x}_i\Tra\V{\beta}))$.

\subsection{Estimation in Low Dimensions}\label{sec:estimation_LD}
\begin{figure}
	\centering
	\includegraphics[scale=0.75]{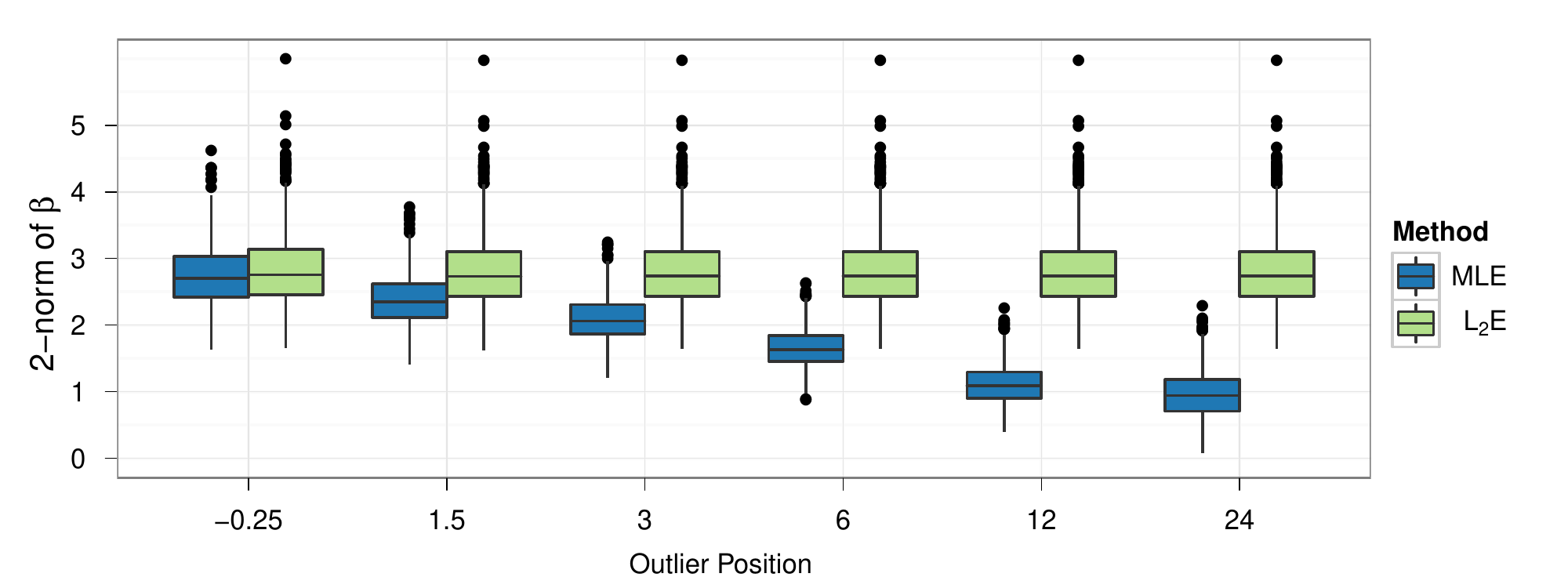}
	\caption[The 2-norm of the regression coefficients as a function of the position of the single outlier.]{The 2-norm of the regression coefficients (intercept not included) as a function of the position of the single outlier.\label{fig:normVshift}}
\end{figure}

\begin{figure}
	\centering
	\includegraphics[scale=0.75]{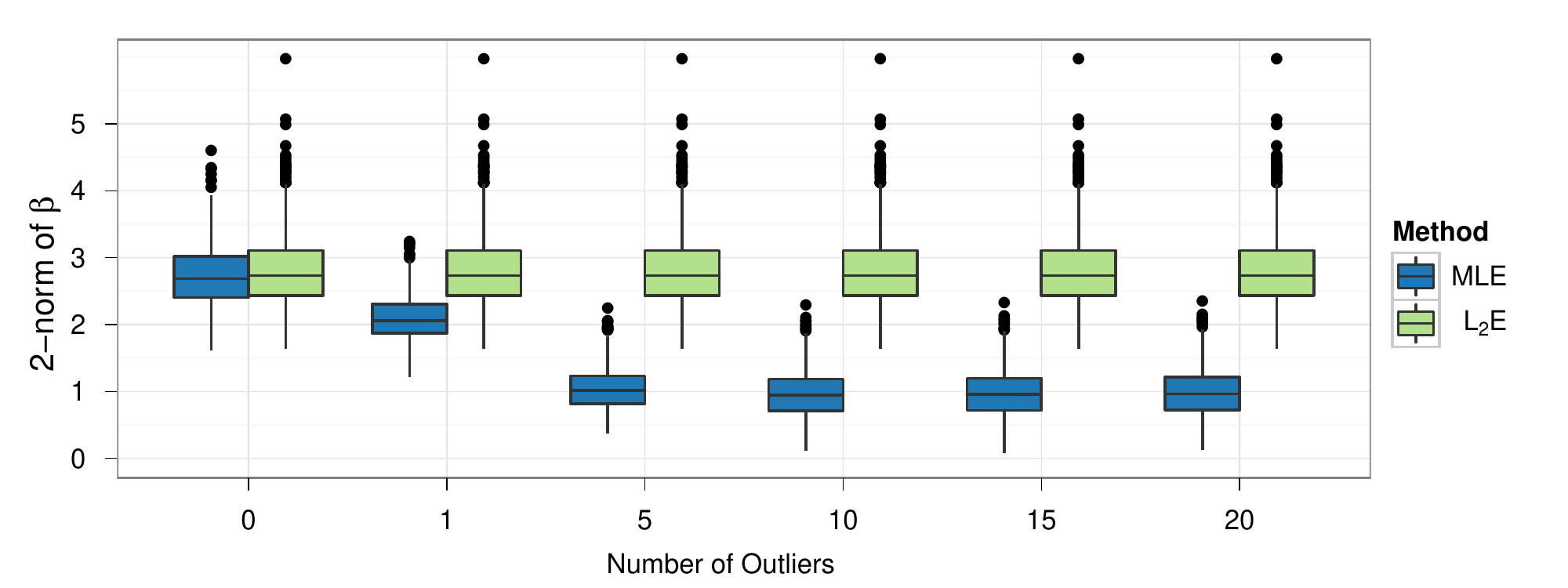}
	\caption[The 2-norm of the regression coefficients as a function of the number of outliers at a fixed position.]{The 2-norm of the regression coefficients (intercept not included) as a function of the number of outliers at a fixed position.\label{fig:normVnumOutliers}}
\end{figure}

In the first scenario, we added a single outlier, $(\VE{Y}{201},\mathbf{x}_{201})$ where $\VE{y}{201} = 0$ and $\V{x}_{201} = (\delta, \delta, \delta, \delta)\Tra$ and $\delta$ took on values in $\{-0.25, 1.5, 3,6, 12, 24\}$. In words, the $201$st point was moved in covariate space along the line that runs through the centroids of the two subpopulations. In the second scenario, we added a variable number of outliers at a single location: $\{(y_i,\V{x}_{i})\}_{i=201}^N$, where $\VE{y}{i} = 0$ and $\V{x}_{i} = (3, 3, 3, 3)\Tra$ for $i = 201, \ldots, N$ and the number of outliers is $N = 0, 1, 5, 10, 15, 20$. For each sequence of scenarios described, we performed logistic regression and L$_2$E regression. Figures~\ref{fig:normVshift} and \ref{fig:normVnumOutliers} summarize the results of first and second scenario, respectively.

The results show two features of the L$_2$E versus the MLE. Consider the first scenario. Figure~\ref{fig:normVshift} shows how $\lVert\Vhat{\beta}\rVert_2$ under each estimation procedure varies with the position of outlier is moved. The MLE values suffer from implosion breakdown as the $201$st point is moved from $-0.25$ to $24$, i.e.\@, $\lVert\Vhat{\beta}\rVert_2$ tends towards 0 as the leverage of the $201$st point increases. In contrast, the L$_2$E is insensitive to the placement of the $201$st point. The second observation is that the L$_2$E's unbiasedness comes at the cost of increased variance. The L$_2$E's spread is greater than the MLE's for all locations of the outlier. Similar behavior is observed in the second scenario. Figure~\ref{fig:normVnumOutliers} shows that implosion breakdown ensues as outliers are added at fixed position. Detailed numerical summaries of the fitted coefficients (sample mean, standard deviation, estimated mean squared error) of these experiments can be found in the Supplementary Materials.

\subsection{Variable Selection in High Dimensions}
\label{sec:var_sel}
In the variable selection experiment we considered a high dimensional variation on the first scenario.  We generated $10$ data sets each with $n = 500$ observations. 
The covariates were drawn from one of three multivariate normal populations. For $i = 1, \ldots 200$, the $\V{x}_i$
are i.i.d.\@ samples from $N(\V{\mu}, 0.75\, \M{I}_p)$. For $i = 201, \ldots, 400$, the $\V{x}_i$ are i.i.d.\@ samples from
$N(-\V{\mu}, 0.75\, \M{I}_p)$. For $i = 401, \ldots, 500$, the $\V{x}_i$ are i.i.d.\@ samples from $N(\V{\nu}, 0.25\, \M{I}_p)$
where $p = 500$, $\VE{\mu}{i} = 0.3$ for $i = 1, \ldots, 50$ and $\VE{\mu}{i} = 0$ for $i = 51, \ldots, 500$,
and $\VE{\nu}{i} = 1$ for $i = 1, \ldots, 50$ and $\VE{\nu}{i} = 0$ for $i = 51, \ldots, 500$.
For $i = 1, \ldots, 400$, the responses were generated independently as $y_i \sim \textsc{B}(1,F(\V{x}_i\Tra\V{\beta}))$, 
where $\beta_0 = 0$ and $\V{\beta} \in \Real^{500}$ with $\beta_i = 1$ for $i = 1, \ldots 50$ and $\beta_i = 0$ for
$i = 51, \ldots, 500$. For $i = 401, \ldots, 500$, the responses were set to $y_i = 0$,

We then performed Elastic Net penalized regression ($\alpha = 0.6$) with the MLE and L$_2$E. Before continuing we note that there are two practical issues that need to be addressed, namely how to choose initial starting points 
since the optimization problem is not convex and how to choose the amount of penalization. In the Supplementary Materials, we describe in detail a heuristic for choosing the initial starting point based on the Karush-Kuhn-Tucker conditions of the optimization problem as well as a robust cross validation scheme for choosing the regularization parameter $\lambda$. To perform the Elastic Net penalized logistic regression we used the {\bf glmnet} package in R \citep{Friedman2010}. We also compared the robust classifier of  \citet{Wang2008} - the Hybrid Huberized Support Vector Machine (HHSVM)  using an MM algorithm. \citet{Wang2008} provide details of the implementation and code for computing the solution paths of the HHSVM. However, their algorithm calculates the paths for a varying LASSO regularization parameter with a fixed ridge regularization parameter because they can be computed quickly by exploiting the piece-wise linearity of the paths under that parameterization of the Elastic Net. Our HHSVM implementation calculates regularization paths using the Elastic Net parameterization used in this article. Details on our implementation can be found in the Supplementary Materials.

\begin{table}[h!]
\caption[Variable Selection Experiment]{True positive count with $n = p = 500$ and 50 nonzero covariates. L$_2$E is the most sensitive method. HHSVM is the least sensitive method. \label{tab:varSelTP}}
\centering
\begin{tabular}{lrrrrrrrrrr} \\
\toprule\midrule
							& \multicolumn{10}{c}{Replicate} \\
		& 1 & 2 & 3 & 4 & 5 & 6 & 7 & 8 & 9 & 10 \\
\midrule
\multirow{1}*{MLE}				& 14 & 10 & 8 & 10 & 1 & 10 & 0 & 14 & 11 & 15 \\		
\multirow{1}*{HHSVM}		 	& 1& 3 & 2 & 2 & 1 & 2 & 1 & 2 & 4 & 2 \\
\multirow{1}*{L$_2$E}		 	& 48 & 47 & 48 & 49 & 48 & 48 & 49 & 46 & 48 & 49\\ 

\bottomrule
\end{tabular}
\end{table}

\begin{table}[h!]
\caption[Variable Selection Experiment]{False positive count with $n = p = 500$ and 50 nonzero covariates. L$_2$E is the most specific method. MLE is the least specific method. \label{tab:varSelFP}}
\centering
\begin{tabular}{lrrrrrrrrrrrr}\\
\toprule\midrule
							& \multicolumn{10}{c}{Replicate} \\
			& 1 & 2 & 3 & 4 & 5 & 6 & 7 & 8 & 9 & 10 \\
\midrule
\multirow{1}*{MLE}	& 141 & 95 & 56 & 148 & 0 & 141 & 0 & 128 & 136 & 170 \\		
\multirow{1}*{HHSVM} & 0 & 4 & 1 & 1 & 1 & 0 & 1 & 0 & 0 & 0 \\	
\multirow{1}*{L$_2$E} & 0 & 0 & 2 & 0 & 0 & 0 & 1 & 1 & 0 & 1 \\

\bottomrule
\end{tabular}
\end{table}

Tables~\ref{tab:varSelTP} and \ref{tab:varSelFP} show the number of true positives and false positives respectively for each method. We see that in scenarios of heavy contamination the  L$_2$E  demonstrates superior sensitivity and specificity compared to both the MLE and HHSVM. It is interesting to note that the MLE tends to be more sensitive than the HHSVM, but at a cost of being drastically less specific. For a closer look comparing the three methods, the cross-validation curves and regularization paths for a replicate can be found in the Supplementary Materials.

\section{Real data examples}
\label{sec:real_data}
\subsection{An  $n > p$ example: Predicting abnormal and normal vertebral columns}

We first consider a real data set in the $n > p$ regime. We present results on the vertebral column data set from the UCI machine learning repository, as described by \cite{Frank+Asuncion:2010}. The data set consists of 310 patients which have been classified as belonging to one of three groups: Normal (100 patients), Disk Hernia (60 patients), Spondylolisthesis (150 patients). In addition to a classification label, six predictor variables are recorded for each patient: pelvic incidence (PI), pelvic tilt (PT), lumbar lordosis angle (LLA), sacral slope (SS), pelvic radius (PR) and grade of spondylolisthesis (GS). All six predictor variables are continuous valued.

% latex table generated in R 2.12.1 by xtable 1.5-6 package
% Fri Sep 16 21:58:42 2011
\begin{table}[ht]
\caption{Correlations among the six biomechanical attributes in the vertebrae data set.}
\label{tab:vertebrae_cor}
\begin{center}
\begin{tabular}{rrrrrrr}
  \hline \hline
 & PI & PT & LLA & SS & PR & GS \\ 
  \hline
PI & 1.00 & 0.63 & 0.72 & 0.81 & -0.25 & 0.64 \\ 
  PT & -- & 1.00 & 0.43 & 0.06 & 0.03 & 0.40 \\ 
  LLA & -- & -- & 1.00 & 0.60 & -0.08 & 0.53 \\ 
  SS & -- & -- & -- & 1.00 & -0.34 & 0.52 \\ 
  PR & -- & -- & -- & -- & 1.00 & -0.03 \\ 
  GS & -- & -- & -- & -- & -- & 1.00 \\ 
   \hline
\end{tabular}
\end{center}
\end{table}

\begin{figure}
	\centering
	\includegraphics[scale=0.75]{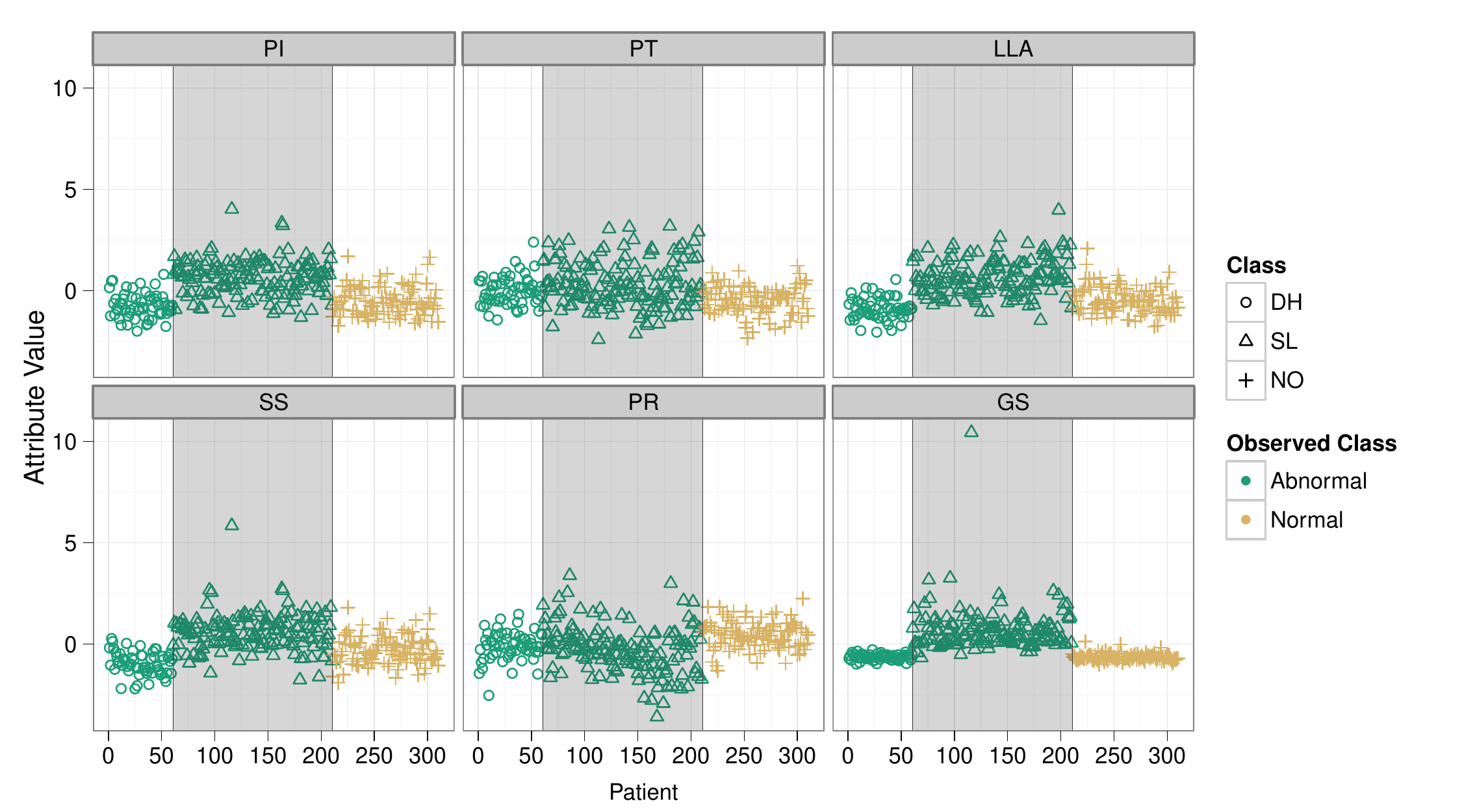}
	\caption{Dot plots of biomechanical attribute values for patients belonging to one of three classes. Patients are randomly ordered within their classes. The attributes are pelvic incidence (PI), pelvic tilt (PT), lumbar lordosis angle (LLA), sacral slope (SS), pelvic radius (PR) and grade of spondylolisthesis (GS). The three underlying classes are Disk Hernia (DH), Spondylolisthesis (SL), and Normal (NO). DH and SL are lumped into the observed class Abnormal. Patients with SL (61 to 210) occupy the plot within the lightly shaded band. \label{fig:vertebrae}}
\end{figure}

\begin{figure}
	\centering
	\includegraphics[scale=0.65]{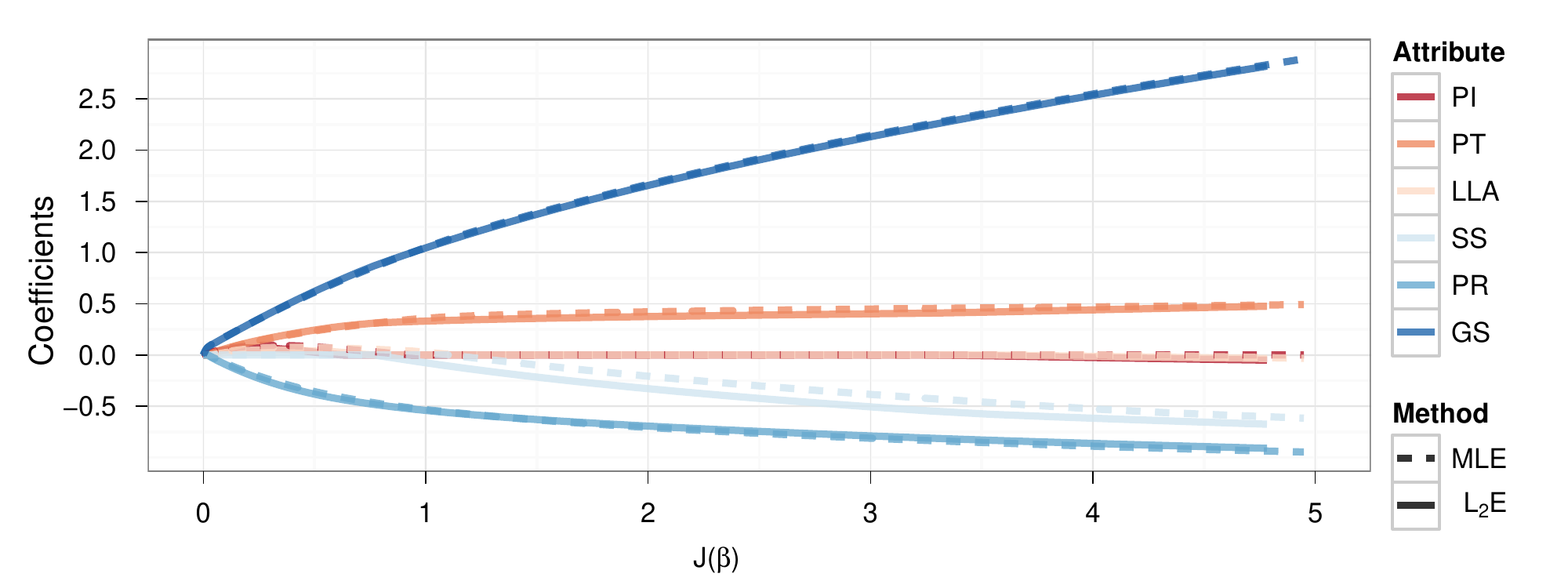}
	\caption{The regularization ($\alpha = 0.2$) paths for the MLE and L$_2$E are very similar for the six biomechanical attributes in the vertebrae data set.\label{fig:vertebrae_regularization_paths}}
\end{figure}

We consider the two class problem of discriminating normal vertebral columns from abnormal ones (Disk Hernia and Spondylolisthesis). Figure~\ref{fig:vertebrae} plots the values of individual covariates for each patient. Table~\ref{tab:vertebrae_cor} shows the correlations between pairs of attributes. Note that the attributes for Disk Hernia and Normal patients overlap a good deal. We may expect similar results as seen in the second simulation scenario described in Section~\ref{sec:estimation_LD} where Disk Hernia patients play the role of a cluster of outlying observations. Due to the correlation, however, the outlying observations are not as distinctly outlying as seen in the simulation examples of Section~\ref{sec:estimation_LD}. Consequently, it also might be anticipated that there will not be differences between the MLE and L$_2$E regularization paths. Indeed, Figure~\ref{fig:vertebrae_regularization_paths} shows the resulting regularization paths generated by the MLE and logistic L$_2$E for $\alpha = 0.2$. The paths are very similar for both methods for other values of $\alpha$ and are not shown. Different initial starting points did not change the resulting logistic L$_2$E regularization paths.

\subsection{An  $n \ll p$ example: A genome wide association study\label{sec:gwa}}

We examine the lung cancer data of \citet{Amos2008}. The purpose of this genome wide association study was to identify risk variants for lung cancer. The authors employed a two stage study using 315,450 tagging SNPs in 1,154 current and former (ever) smokers of European ancestry and 1,137 frequency matched, ever-smoking controls from Houston, Texas in the discovery stage. The most significant SNPs found in the discovery phases were then tested in a larger replication set. Two SNPs, rs1051730 and rs8034191, on chromosome 15 were found to be significantly associated with lung cancer risk in the validation set. SNP markers can have a high degree of collinearity due to recombination mechanics. SNPs that are physically close to each other tend to be highly correlated and are said to be in linkage disequilibrium. The pair rs1051730 and rs8034191 for example are in ``high" linkage disequilibrium. 

In this section we reexamine the discovery data using logistic L$_2$E and the logistic MLE. Note that it is current practice of geneticists to do univariate inference with an adjustment for multiple testing and this approach was taken in \citet{Amos2008}. Taking a multivariate approach as will be done in this section, however, allows the analyst to take into account dependencies between the SNPs. As an initial comparison we consider a subset of the entire data set and restrict our analysis to SNPs on chromosome 15. We impute missing genotypes at a SNP by using the MACH 1.0 package, a Markov Chain based haplotyper \citep*{Li2006}. After missing data are imputed and keeping only imputations with a quality score of at least 0.9, 8,701 SNPs are retained on 1152 cases and 1136 controls.

\begin{figure}
\centering
	\includegraphics[scale=.625]{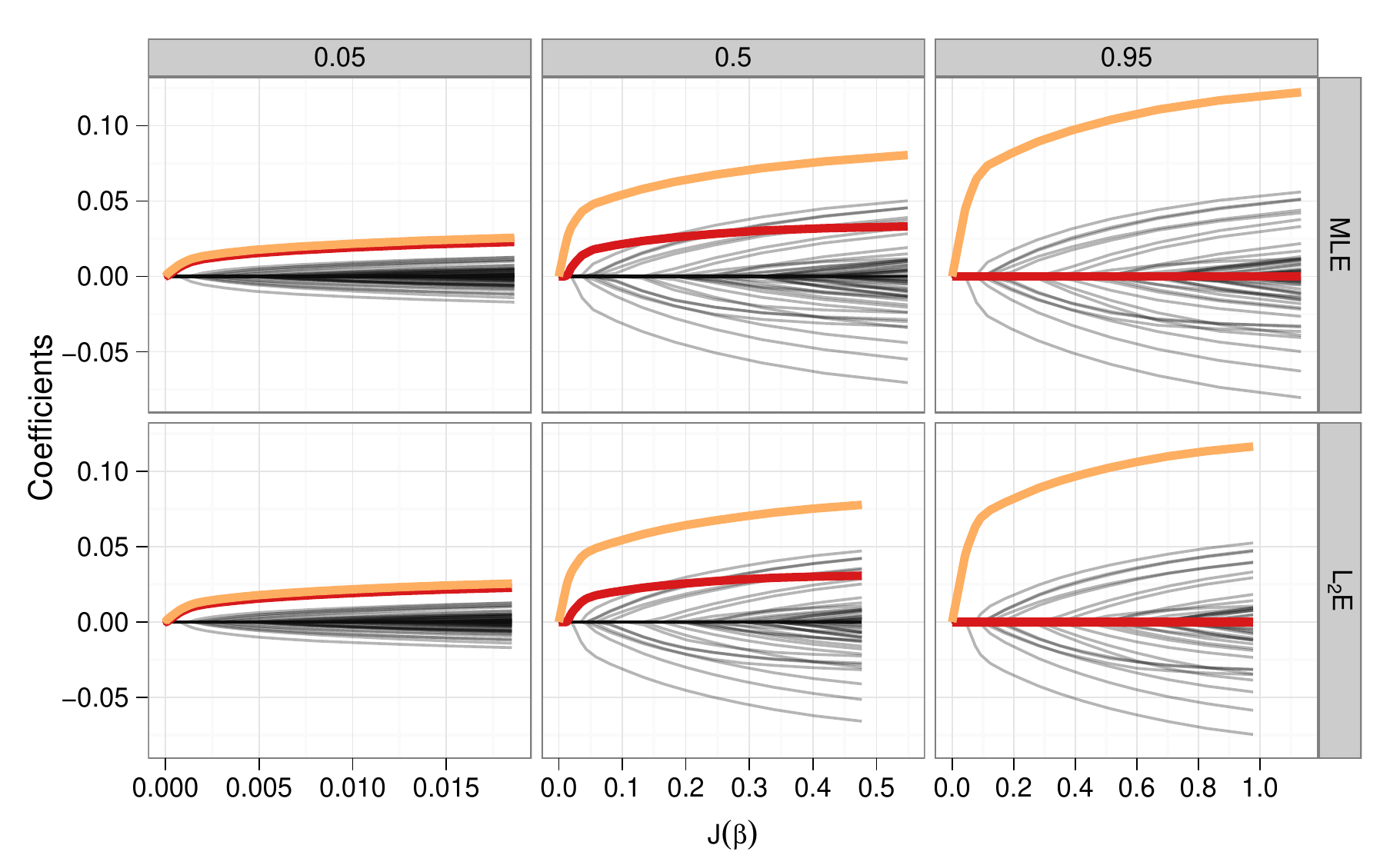}
\caption{Regularization paths of regression coefficients of SNP markers on Chromosome 15 for L$_2$E and MLE for $\alpha = 0.05, 0.5$, and $0.95$. The regularization paths for rs1051730 are in light-thick lines; the paths for rs8034191 are in dark-thick lines. The L$_2$E and MLE paths are nearly identical. For $\alpha =0.95$, i.e.\@ nearly LASSO regression, rs8034191 was not selected for the shown range of penalizations by either method.\label{fig:regPath}}
\end{figure}

Figure~\ref{fig:regPath} summarizes the variable selection results for the logistic L$_2$E and MLE for $\alpha = 0.05, 0.5,$ and $0.95$. There are three things to note. First, the regularization paths for the L$_2$E and MLE are almost identical. Second, both methods produce regularization paths that identify rs1051730 (light-thick line) and rs8034191 (dark-thick line) as having the greatest partial correlation with the case/control status. Third, the paths for rs1051730 and rs8034191 behave as would be expected with $\alpha$. For small $\alpha$, or more ridge-like penalty, the two paths become more similar. For large $\alpha$, or more LASSO-like penalty,
only one of the two correlated predictors enters the model while the other is excluded.

\section{Discussion}
\label{sec:discussion}
Outliers can introduce bias in some commonly used maximum likelihood estimation procedures. This well known fact, however, warrants attention because bias can have material effects on the ubiquitous LASSO-based variable selection procedures. In the context of standard logistic regression, influential outliers cause implosion breakdown. In this paper we have demonstrated that the combination of implosion breakdown and the soft-thresholding mechanism of LASSO variable selection can lead to missed detection of relevant predictors.

To guard against the undue influence of outliers on estimation and variable selection for binary responses, we propose a robust method for performing sparse logistic regression. Our method is based on minimizing the estimated L$_2$ distance between the logistic parametric model and the underlying true conditional distribution. The resulting optimization problem is a penalized non-linear least squares problem which we solve with an MM algorithm. Our MM algorithm in turn reduces the optimization problem to solving a series penalized least squares problems whose solution paths can be solved very efficiently with coordinate descent and warm starts.

Although we present our work as a method for robust binary logistic regression, our method immediately extends to other related contexts. Our algorithm can be extended to handle more than two classes. The generalization to the $K$-class multinomial is straightforward.
\begin{equation*}
	L(\M{Y}, \Mtilde{X}\M{\Theta}) = \sum_{k=1}^K \lVert \V{y}_k - F_k(\Mtilde{X}\V{\Theta}) \rVert_2^2, 
\end{equation*}
where $\ME{Y}{ik} = 1$ if the $i$th observation belongs to class $k$ and 0 otherwise and the $i$th element of vector $F_k(\Mtilde{X}\M{\Theta})$ is given by
\begin{equation*}
	\frac{\exp(\Vtilde{x}_i\Tra\V{\theta}_k)}{1 + \sum_{j=1}^K \exp(\Vtilde{x}_i\Tra\V{\theta}_j)}.
\end{equation*}
This non-linear least squares problem also has bounded curvature and consequently can also be solved by minimizing a sequence of LASSO-penalized least squares problems.

Our algorithm can also be used as a subroutine in performing robust binary principal component analysis and, more generally, robust binary tensor decompositions. A common strategy in array decompositions for multiway data, including multiway binary data, is to use block coordinate descent or alternating minimization \citep*{ Collins2001, Kolda2009, Lee2010}. For binary multiway data, each block minimization would perform a batch of independent robust logistic regressions.

We want to make clear that the logistic L$_2$E is not a competitor to the MLE but rather a complement. Both methods are computationally feasible and can be run on data together. As seen in the real data examples of Section~\ref{sec:real_data}, sometimes the logistic L$_2$E recovers the MLE solution. On the other hand, when discrepancies do occur, taking the MLE and L$_2$E solutions together can provide insight into the data that would be harder to identify with the MLE solution alone.

We close with some interesting directions for future work. We have seen that LASSO-based variable selection in the presence of implosion breakdown can lead to missed detection of relevant predictors. This motivates the question of whether explosion breakdown can lead to the inclusion of irrelevant predictors. Finally, with respect to convergence issues of our algorithm, while we have established conditions under which our algorithm is guaranteed to converge to a stationary point we do not have rigorous results on the rate at which it does so. As a complement to methods that may be sensitive to the presence of outliers, characterizing the convergence speed of our algorithm has a great deal of practical importance.

\bigskip
\begin{center}
{\large\bf SUPPLEMENTAL MATERIALS}
\end{center}

\begin{description}

\item[Algorithm details, simulation results, proofs, and derivations:] The Supplementary Materials includes additional details on the algorithm (e.g.\@ choosing initial starting points, stopping criteria, and choosing regularization parameters), additional results from the estimation experiments in Section~\ref{sec:estimation_LD} and variable selection experiments in Section~\ref{sec:var_sel}, proofs for Theorems~\ref{thm:MM} and \ref{thm:convergence_n_g_p}, and a derivation of our HHSVM algorithm. (Supplement.pdf)

\item[Code:] C and R code used to generate results shown in the article along with relevant data have also been made available. A readme file details how to compile and run the code. The SNP data is not included for confidentiality reasons. (GNU zipped tar file) 

\end{description}

\begin{center}
{\large\bf ACKNOWLEDGMENTS}
\end{center}
The authors thank Christopher Amos for generously allowing them to work with the lung cancer data set. All plots were made
using the open source R package ggplot2 \citep{Wickham2009}. Eric Chi was supported by grant DE-FG02-97ER25308 from the Department of Energy. David Scott was supported in part by grant DMS-09-07491 from the National Science Foundation. 

\section*{Supplementary Materials}

\section{Proofs}\label{sec:proofs}

\subsection{Proof of Theorem 4.1}\label{app:MMproof}

It is immediate that $L(\Vtilde{\theta}; \Vtilde{\theta}) = L(\V{y}, \Mtilde{X}\Vtilde{\theta})$. We turn our attention to proving that $L(\V{\theta}; \Vtilde{\theta}) \geq L(\V{y}, \Mtilde{X}\V{\theta})$ for all $\V{\theta}, \Vtilde{\theta} \in \Real^{p+1}$. Since $L(\V{y}, \Mtilde{X}\V{\theta})$ has bounded curvature our strategy is to represent $L(\V{y}, \Mtilde{X}\V{\theta})$ by its exact second order Taylor expansion about $\Vtilde{\theta}$ and then find a tight uniform bound over the quadratic term in the expansion. This approach applies in general to functions with continuous second derivative and bounded curvature \citep{Boehning1988}.

The exact second order Taylor expansion of $L(\V{y}, \Mtilde{X}\V{\theta})$ at $\Vtilde{\theta}$ is given by
\begin{equation*}
	L(\V{y}, \Mtilde{X}\V{\theta}) = L(\V{y}, \Mtilde{X}\Vtilde{\theta}) + (\V{\theta} -\Vtilde{\theta})\Tra \nabla L(\V{y}, \Mtilde{X}\V{\theta}) + \frac{1}{2}(\V{\theta} -\Vtilde{\theta})\Tra \M{H}_{\V{\theta}^*} (\V{\theta}-\Vtilde{\theta}),
\end{equation*}
where $\V{\theta}^* = \gamma \Vtilde{\theta} + (1-\gamma) \V{\theta}$ for some $\gamma\in (0, 1)$ and
\begin{equation*}
\begin{split}
	\nabla L(\V{y}, \Mtilde{X}\V{\theta}) &= 4 n^{-1} \M{X}\Tra \M{G} (\V{p} -  \V{y}) \\
	\M{H}_\V{\theta} &= \frac{2}{n} \M X \Tra \M M_\V{\theta} \M X, \\
	\M{G} &= \operatorname{diag}\{\VE{p}{1}(1-\VE{p}{1}), \ldots, \VE{p}{n}(1-\VE{p}{n})\} \\
	\M{M}_\V{\theta} &= \operatorname{diag}\{\psi_{\VE{U}{1}}(\VE{P}{1}), \ldots, \psi_{\VE{U}{n}}(\VE{P}{n})\} \\
	\V{u} &= 2\V{y} -\V{1} \\
	\V{p} &= F(\Mtilde{X}\V{\theta}) \\
	\psi_u(p) &=  [2p(1-p) - (2p-1) ((2p-1) - u)]p(1-p). \\
\end{split}
\end{equation*}

Note that $(\M{M}_\V{\theta})_{ii}$ is bounded from above, i.e.\@, $\sup_{\V{\theta}\in\Theta} (\M{M}_\V{\theta})_{ii} < \infty$. We now introduce a surrogate function:
 \begin{equation*}
	L(\V{\theta}; \Vtilde{\theta}) = L(\V{y}, \Mtilde{X}\Vtilde{\theta}) + \frac{4}{n} (\V{\theta}-\Vtilde{\theta})\Tra \M X\Tra \M G (F(\Mtilde{X}\Vtilde{\theta})- \V{y}) +  \frac{\eta}{n}(\V{\theta} - \Vtilde{\theta})\Tra \M X \Tra \M X (\V{\theta}-\Vtilde{\theta}),
\end{equation*}
where 
\begin{equation*}
	\eta \geq \max \left\{ \sup_{p \in [0,1]} \psi_{-1}(p), \sup_{p \in [0,1]} \psi_{1}(p) \right \}.
\end{equation*}
Note that for any $\V{\theta} \in \Real^{p+1}$, $(\M{M}_\V{\theta})_{ii} \leq \eta$. Therefore,
\begin{equation*}
	\begin{split}
	(\V{\theta} -\Vtilde{\theta})\Tra \M{H}_{\V{\theta}^*} (\V{\theta}-\Vtilde{\theta}) &= 
	(\V{\theta} -\Vtilde{\theta})\Tra \M X\Tra \M M_{\V{\theta}^*}\M X (\V{\theta}-\Vtilde{\theta}) \\
	&\leq \eta (\V{\theta}-\Vtilde{\theta})\Tra \M X\Tra \M X (\V{\theta} -\Vtilde{\theta}), \\
	\end{split}
\end{equation*}
and consequently $L(\V{\theta}; \Vtilde{\theta})$ majorizes $L(\V{y}, \Mtilde{X}\Vtilde{\theta})$ at $\Vtilde{\theta}$.
\qed

The following observations lead to a simpler lower bound on $\eta$. Note that
\begin{equation*}
	\sup_{p \in [0,1]} \psi_{-1}(p) = \sup_{p \in [0,1]} \psi_1(p),
\end{equation*}
since $\psi_{-1}(p) = \psi_1(1-p)$. So, the lower bound on $\eta$ can be more simply expressed as
\begin{equation}
	\label{eq:bound}
	\sup_{p \in [0,1]} \psi_1(p) = \max_{p \in [0,1]} \psi_1(p) =  \frac{1}{4}\max_{q \in [-1,1]} \left \{ \frac{3}{2} q^4 - q^3 - 2 q^2 + q + \frac{1}{2} \right \}. \\
\end{equation}
The first equality follows from the compactness of $[0,1]$ and the continuity of $\psi_1(p)$. The second equality follows from reparameterizing $\psi_1(p)$ in terms of $q=2p-1$. Since the derivative of the polynomial in (\ref{eq:bound}) has a root at $1$, it is straightforward to argue that the lower bound of $\eta$ is attained at the second largest root, which is $(-3 + \sqrt{33})/12$. Thus, the majorization holds so long as

\begin{equation*}
	\eta \geq \frac{3}{16} q^4 - \frac{1}{4}q^3 - \frac{1}{2}q^2 + \frac{1}{4}q + \frac{1}{16} \Bigg |_{q = \frac{-3 + \sqrt{33}}{12}}. \\
\end{equation*}

\subsection{Proof of Theorem 4.2}\label{sec:GCproof}

A key condition in MM algorithm convergence proofs is coerciveness since it is a sufficient condition to ensure the existence of a global minimum. Recall that 
a continuous function $f: U \subset \Real^n \rightarrow \Real$ is coercive if all its level sets $S_t = \{ \V{x} \in U :  f(\V{x}) \leq t\}$ are compact. 

We will use the MM algorithm global convergence results in \cite{Schifano2010}. Let $\xi(\V{\theta})$ denote the objective function and let $\xi^{[S]}(\V{\theta}, \Vtilde{\theta})$ denote a surrogate objective function that will be minimized with respect to its first argument in lieu of $\xi(\V{\theta})$. The iteration map $\varphi$ is given by
\begin{equation*}
	\varphi(\Vtilde{\theta}) = \underset{\V{\theta}}{\arg\min}\, \xi^{[S]}(\V{\theta}, \Vtilde{\theta}).
\end{equation*}
We now state a slightly less general set of regularity conditions than those in \cite{Schifano2010} that are sufficient for our purposes. Suppose $\xi, \xi^{[S]},$ and $\varphi$ satisfy the following set of conditions:
\begin{itemize}
	\item[R1.] The objective function $\xi(\V{\theta})$ is locally Lipschitz continuous for $\V{\theta} \in \Theta$ and coercive. The set of stationary points $\mathcal{S}$ of 
	$\xi(\V{\theta})$ is a finite set, where the notion of a stationary point is defined as in \cite{Clarke1983}.
	\item[R2.] $\xi(\V{\theta}) = \xi^{[S]}(\V{\theta},\V{\theta})$ for all $\V{\theta} \in \Theta.$
	\item[R3.] $\xi^{[S]}(\V{\theta},\Vtilde{\theta}) < \xi^{[S]}(\V{\theta},\V{\theta})$ for all $\V{\theta}, \Vtilde{\theta} \in \Theta$ where $\V{\theta} \not = \Vtilde{\theta}$.
	\item[R4.] $\xi^{[S]}(\V{\theta},\Vtilde{\theta})$ is continuous for $(\V{\theta}, \Vtilde{\theta}) \in \Theta\times\Theta$ and locally Lipschitz in $\Theta$.
	\item[R5.] $\varphi(\V{\theta})$ is a singleton set consisting of one bounded vector for $\V{\theta} \in \Theta$.
\end{itemize}
Then $\{\Vn{\theta}{n}, n \geq 0\}$ converges to a fixed point of the iteration map $\varphi$. By Proposition A.8 in \cite{Schifano2010} the fixed points of $\varphi$ coincide with $\mathcal{S}$.

In our case we have the following objective and surrogate functions
\begin{equation*}
\begin{split}
	\xi(\V{\theta}) &= \frac{1}{2n}\lVert \V{y} - F(\Mtilde{X}\V{\theta})\rVert_2^2 + \lambda \left (\alpha \lVert \V{\beta} \rVert_1+ \frac{(1-\alpha)}{2}\lVert \V{\beta} \rVert_2^2 \right ) \\
	\xi^{[S]}(\V{\theta}, \Vtilde{\theta}) &= \frac{1}{2}L(\V{\theta}, \Vtilde{\theta}) + \lambda \left (\alpha \lVert \V{\beta} \rVert_1+ \frac{(1-\alpha)}{2}\lVert \V{\beta} \rVert_2^2 \right ). \\
\end{split}
\end{equation*}
We check each regularity condition in turn.

\begin{itemize}
	\item[R1.] Since $\lVert \V{y} - F(\Mtilde{X}\V{\theta})\rVert_2^2$ is bounded below and the penalty term is coercive, $\xi(\V{\theta})$ is coercive. Recall that the gradient of the $L(\V{y}, \Mtilde{X}\V{\theta})$ is $(4/n) \M X\Tra \M G (F(\Mtilde{X}\V{\theta}) - \V{y})$. The norm of the gradient is bounded; specifically it is no greater than $2\sigma_1^2$ where $\sigma_1$ is the largest singular value of $\M{X}$. Therefore, $L(\V{y}, \Mtilde{X}\V{\theta})$ is Lipschitz continuous and therefore locally Lipschitz continuous. Consequently, $\xi(\V{\theta})$ is locally Lipschitz continuous. If the set of stationary points of $\xi(\V{\theta})$ is finite, then R1 is met.
	\item[R2 and R3.] Recall the majorization we are using is given by
	 \begin{equation*}
		L(\V{\theta}; \Vtilde{\theta}) = L(\V{y}, \Mtilde{X}\Vtilde{\theta}) + (\V{\theta}-\Vtilde{\theta})\Tra \nabla L(\V{y}, \Mtilde{X}\Vtilde{\theta})  +  \frac{\eta}{n}(\V{\theta} - 	\Vtilde{\theta})\Tra \M X \Tra \M X (\V{\theta}-\Vtilde{\theta}),
	\end{equation*}
	where
	\begin{equation*}
		\eta > \frac{1}{4}\max_{q \in [-1,1]} \left \{ \frac{3}{2} q^4 - q^3 - 2 q^2 + q + \frac{1}{2} \right \}. \\
	\end{equation*}
	To ensure that the majorization is strict we need the inequality to be strict. Thus, the curvature of the majorization exceeds the maximum curvature of $L(\V{y},\Mtilde	{X}\V{\theta})$ and the majorization is strict. R2 and R3 are met.
	\item[R4.] The penalized majorization is the sum of continuous functions in $(\V{\theta},\Vtilde{\theta}) \in \Theta \times \Theta$ and is consequently continuous. The penalized majorization as a function of its first argument is the sum of  a positive definite quadratic
function and the 1-norm function, both of which are locally Lipschitz continuous so their sum is locally Lipschitz continuous. R4 is met.
	\item[R5.] If $\lambda(1-\alpha) > 0$ then $\xi^{[S]}(\V{\theta}, \Vtilde{\theta})$ is strictly convex in $\V{\theta}$ and thus has at most one global minimizer. Since 
$\xi^{[S]}(\V{\theta}, \Vtilde{\theta})$ is also coercive in $\V{\theta}$ it has at least one global minimizer. R5 is met.
\end{itemize}
Thus, Algorithm 1 will converge to a stationary point of  $\xi(\V{\theta})$, provided that there are only finitely many stationary points and the coordinate descent minimization of the Elastic Net penalized quadratic majorization is solved exactly. \qed

\begin{remark}
If $\xi$ does not have finitely many stationary points, it can be shown that the limit points of the sequence of iterates are stationary points and that the set of limit points is connected (\citealp{Schifano2010}; \citealp{Chi2011}).
\end{remark}

\begin{remark}
The iterate update $\Vn{\theta}{m+1} = \varphi(\Vn{\theta}{m})$ can be accomplished by any means algorithmically so long as the global minimum of the majorization is found. Iterates of coordinate descent are guaranteed to converge to a global
minimizer provided that the loss is differentiable and convex and the penalty is convex and separable \citep{Tseng2001}. Thus, applying coordinate descent on the Elastic Net penalized quadratic majorization will find the global minimum.
\end{remark}

\begin{remark}
Our definition of stationary points has to change because the objective functions of interest are locally Lipschitz continuous and therefore differentiable almost everywhere except on a set of Lebesgue measure zero. \cite{Clarke1983} defines and proves properties of a generalized gradient for locally Lipschitz functions. Apart from pathological cases, when a function is convex the generalized gradient is the subdifferential. See Proposition 2.2.7 in \cite{Clarke1983}.
When a function is differentiable the generalized gradient is the gradient. Thus as would be expected a point $\V{x}$ is a stationary point of a locally Lipschitz function if the function's generalized gradient at $\V{x}$ contains $\V{0}$.
\end{remark}

\section{Algorithm Details}\label{sec:algo}
Algorithm~\ref{alg:iterSolve} gives pseudocode for the resulting iterative solver for a given pair of parameters $\alpha$ and $\lambda$. The symbol $*$ denotes the Hadamard element-wise product. In practice we also use active sets
to speed up computations. That is, for a given initial $\V{\beta}$, we only update the non-zero coordinates of $\V{\beta}$, the active set, until there is little change in the active set parameter estimates. The non-active set parameter estimates are then updated once. If they remain zero, the Karush-Kuhn-Tucker (KKT) conditions have been met and a global minimum of (4.4) has been found. If not, then the active set is expanded to include the coordinates whose KKT conditions have been violated and the process is repeated.

\begin{algorithm}[t]
	\begin{algorithmic}[0]
		\STATE $\V{\theta} \leftarrow \text{initial guess}$
		\REPEAT
			\STATE $\V{p} \leftarrow F(\Mtilde{X}\V{\theta})$
			\STATE $\M{G} \leftarrow \operatorname{diag}\{\V{p} * (\V{1} - \V{p})\}$
			\STATE $\V{z} \leftarrow 2\M{G}(\V{p} - \V{y})$
			\STATE $\V{\zeta} \leftarrow \M{X}\V{\beta} - \frac{1}{\eta}(\V{z} - \overline{z}\V{1})$
			\STATE $\beta_0 \leftarrow \beta_0 - \eta^{-1}\overline{z}$
			\REPEAT
				\FOR{$k = 1..p$}
					\STATE $\V{r} \leftarrow \V{\zeta} - (\M{X}\V{\beta} - \VE{\beta}{k}\V{x}_k)$
					\STATE $\VE{\beta}{k}  \leftarrow S\left(\frac{\eta}{n}\V{X}_k\Tra\V{r},\lambda \alpha\right )\big /\left [\frac{\eta}{n} \lVert \V{X}_k\rVert_2^2 + \lambda (1-\alpha) \right ]$
				\ENDFOR
			\UNTIL{convergence}
		\UNTIL{convergence}
		\RETURN $\V{\theta}$
	\end{algorithmic}
	\caption{\, \textsc{Iterative L$_2$E solver}}	
	 \label{alg:iterSolve}
\end{algorithm}

\subsection{Choosing the penalty parameters}
\label{sec:model_selection}
\subsubsection{Warm Starts and Calculating Regularization Paths}
We will need to compare the regression coefficients obtained at many values of the penalty parameter $\lambda$ to perform model selection. Typically we can rapidly calculate regression coefficients for a decreasing sequence of values of $\lambda$ through warm starts. Namely, a solution to the problem using $\lambda_k$ as a regularization parameter is used as the initial starting value for the iterative algorithm applied to the subsequent problem using $\lambda_{k+1}$ as a regularization parameter. The idea is if $\lambda_k$ and $\lambda_{k+1}$ are not too far apart, the solutions to their corresponding optimization problems will be close to each other. Thus, the solution of one optimization problem will be a very good initial starting point for the succeeding optimization problem.

For $\lambda$ sufficiently large, only the intercept term $\theta_0$ will come into the model. The smallest $\lambda^*$ such that all regression coefficients are shrunk to zero is given by
\begin{equation}
	\lambda^* = \frac{2}{n\alpha} \overline{y}(1-\overline{y})\max_{j=1,\ldots,p} \lvert \V{x}_{(j)}\Tra\V{y} \rvert, \\
\end{equation}
where $\V{x}_{(j)}$ denotes the $j$th column of the design matrix $\M{X}$. We compute a grid of $\lambda$ values equally spaced on a log scale between $\lambda_{\max} = \lambda^*$ and $\lambda_{\min} = \epsilon \lambda_{\max}$ where $\epsilon < 1$. In practice, we have found the choice of $\epsilon = 0.05$ to be useful. In general, we are not interested in making $\lambda$ so small as to include all variables.

Moreover, due to the possible multi-modality of the L$_2$E loss, we recommend computing the regulation paths starting from a smaller regularization parameter and increasing the parameter value until $\lambda_{\max}$. Since we face multi-modality initial starting points can make a significant difference in the answers obtained.

\subsubsection{The heuristic for choosing starting values}
\label{sec:heuristic}
Since the logistic L$_2$E loss is not convex, it may have multiple local minima. For the purely LASSO-penalized problem, the KKT condition at a local minimum is
\begin{equation*}
	\VE{\nu}{j} = \lvert \V{X}_{(j)}\Tra \M{G} (\V{y} - F(\beta_0\V{1} + \M{X}\V{\beta})) \rvert \leq \lambda.
\end{equation*}
Equality is met whenever $\VE{\beta}{j} \not = 0$. Thus, the largest values of $\VE{\nu}{j}$
will correspond to a set of covariates which include covariates with non-zero regression coefficients. The leap of faith is that the largest values of $\VE{\nu}{j}$
evaluated at the null model will also correspond to a set of covariates which include covariates with non-zero regression coefficients. This idea has been used in a ``swindle" rule \citep{Wu2009} and STRONG rules for discarding variables \citep*{Tibshirani2011}. In those instances the goal is to solve a smaller optimization problem. In contrast, we initialize starting parameter entries to zero rather than excluding variables with low scores from the optimization problem. Specifically, we do the following:
\begin{inparaenum}[(1)]
  \item calculate the following scores $\VE{z}{j} = \lvert \V{X}_{(j)}\Tra \M{G}_0 (\V{y} - p\V{1})) \rvert$,
where $p = \overline{y}$ the sample mean of $\V{y}$ and $\M{G}_0 = p(1-p)\M{I}$;
  \item set $\beta_0^{(0)} = \log(\overline{y} / (1-\overline{y}))$; and
  \item set $\beta_j^{(0)} = I(j \in \mathcal{S})$,
\end{inparaenum}
where $I(\cdot)$ denotes the indicator function and $\mathcal{S} = \{j : \VE{z}{j}$ is ``large"$\}$.

\subsubsection{Robust Cross-Validation}
\label{sec:robust_cross_validation}
Once we have a set of models computed at different regularization parameter values, we select the model that is optimal with respect to some criterion. We use the following robust 10-fold cross-validation scheme to select the model. After partitioning the data into 10 training and test sets, for each $i = 1, \ldots, 10$ folds we compute regression coefficients 
$\Vhat{\theta}^{-i}(\lambda)$ for a sequence of $\lambda$'s between $\lambda_{\max}$ and $\lambda_{\min}$ holding out the $i$th test set $\mathcal{S}_i$. 

Next we refit the model using the reduced variable set $\mathcal{S}^c_i$, those with nonzero regression coefficients, and refit using logistic L$_2$E with $\alpha = 0$. This refitting produces less biased estimates. We are adopting the same strategy as LARS-OLS in \citet*{Efron2004}. Our framework, however, could adopt a more sophisticated strategy along the lines of the Relaxed LASSO in \citet{Meinshausen2007}. Henceforth let 
$\Vhat{\theta}^{-i}(\lambda)$ denote the regression coefficients obtained after the second step. Let $d_j^{-i}(\lambda)$ denote the contribution of observation $j$ to the L$_2$E loss under the model
$\Vhat{\theta}^{-i}(\lambda)$, i.e.\@,
\begin{equation*}
d_{j}^{-i}(\lambda) = \left (\VE{y}{j} - F(\Vtilde{X}_j\Tra \Vhat{\theta}^{-i}(\lambda)) \right )^2.
\end{equation*}

We use the following criterion to choose $\lambda^*$:
\begin{equation*}
	\lambda^* = \underset{\lambda}{\arg\min} \; \left \{\underset{i =1, \ldots, 10}{\operatorname{median}}
	\left\{\underset{j \in \mathcal{S}_i}{\operatorname{median}}\; d_j^{-i}(\lambda) \right \} \right \}.
\end{equation*}

The reason for choosing $\lambda^*$ in this way is due to a feature of the robust fitting procedure. Good robust models will assign unusually large values of $d_j^{-i}(\lambda)$ to outliers. Thus, the total L$_2$E loss
is an inappropriate measure of the prediction error if influential outliers were present. On the other hand, taking the median, for example, would provide a more unbiased measure of the prediction error regardless of outliers. The final model selected would be the one that minimizes the robust prediction error criterion.

\newpage
\section{Simulation Experiments in Low Dimensions}
Tables~\ref{tab:shift} and \ref{tab:nc} provide summary statistics for simulations performed in Section 5.1. The experiments show the unbiasedness of the L$_2$E compared to the MLE at the price of increased variance.
% which suffers implosion breakdown as contamination becomes more extreme.
%The price for unbiasedness is that the L$_2$E sample standard error is greater than that of the MLE for all cases. 
The mse summarizes the bias-variance tradeoff between the two methods.

\setcounter{table}{2} 
\begin{table}[h!]
%The true parameter value is $\V{\theta} = (0, 1, 0.5,1,2)\Tra$.
\caption[Effect of varying the position of a single outlier.]{
Effect of varying the position of a single outlier 
%The L$_2$E estimates  are essentially
%          unbiased regardless of the location of the outlier, while the MLE results become increasingly biased as the outlier is moved 
from 
$-0.25$ to $24$.
           \label{tab:shift}}
\centering
\begin{tabular}{ccccccccc}\\
\toprule
				&		& 	& \multicolumn{3}{c}{MLE} & \multicolumn{3}{c}{L$_2$E} \\
Outlier Position 	& Coefficient	& True Value &	mean 	& std & mse & mean & std & mse\\
\midrule
\multirow{5}*{-0.25}	& $\V{\beta}_0$ & 0	& -0.002 & 0.182 & 0.033 & -0.005 & 0.192 & 0.037 \\ 
				& $\V{\beta}_1$ 	& 1 & 1.032 & 0.434 & 0.189 & 1.063 & 0.480 & 0.234 \\ 
				& $\V{\beta}_2$ 	& 0.5 & 0.526 & 0.424 & 0.180 & 0.539 & 0.463 & 0.216\\ 
				& $\V{\beta}_3$ 	& 1 & 1.047 & 0.439 & 0.195 & 1.079 & 0.482 & 0.238 \\ 
  				& $\V{\beta}_4$ 	& 2 & 2.110 & 0.487 & 0.249 & 2.181 & 0.572 & 0.359 \\ 				
\midrule		
\multirow{5}*{1.5}	& $\V{\beta}_0$ & 0 	& -0.024 & 0.168 & 0.029 & 0.002 & 0.192  & 0.037 \\ 
  				& $\V{\beta}_1$ 	& 1 & 0.868 & 0.394 & 0.173 & 1.052 & 0.476 & 0.229 \\ 
  				& $\V{\beta}_2$ 	& 0.5 & 0.401 & 0.391 & 0.162 & 0.532 & 0.460 & 0.212 \\ 
  				& $\V{\beta}_3$ 	& 1 & 0.880 & 0.396 & 0.171 & 1.068 & 0.478 & 0.233 \\ 
  				& $\V{\beta}_4$ 	& 2 & 1.860 & 0.430 & 0.204 & 2.160 & 0.567 & 0.347\\ 		
\midrule	
\multirow{5}*{3}	  	& $\V{\beta}_0$ & 0	& -0.022 & 0.157 & 0.025 & 0.002 & 0.192 & 0.037 \\ 
  				& $\V{\beta}_1$ 	& 1 & 0.732 & 0.368 & 0.207 & 1.054 & 0.476 & 0.229 \\ 
  				& $\V{\beta}_2$ 	& 0.5 & 0.296 & 0.369 & 0.178 & 0.533 & 0.460 & 0.212 \\ 
  				& $\V{\beta}_3$ 	& 1 & 0.743 & 0.368 & 0.201 & 1.069 & 0.478 & 0.233 \\ 
  				& $\V{\beta}_4$ 	& 2 & 1.662 & 0.392 & 0.268 & 2.163 & 0.567 & 0.347 \\ 
\midrule				
\multirow{5}*{6}	  	& $\V{\beta}_0$ & 0	& -0.020 & 0.142 & 0.021 & 0.002 & 0.192 & 0.037 \\
  				& $\V{\beta}_1$ 	& 1 & 0.508 & 0.337 & 0.356 & 1.054 & 0.476 & 0.229 \\ 
  				& $\V{\beta}_2$ 	& 0.5 & 0.112 & 0.344 & 0.268 & 0.533 & 0.460 & 0.212 \\ 
  				& $\V{\beta}_3$ 	& 1 & 0.516 & 0.334 & 0.346 & 1.069 & 0.478 & 0.233 \\ 
  				& $\V{\beta}_4$ 	& 2 & 1.350 & 0.347 & 0.543 & 2.163 & 0.567 & 0.347 \\  
\midrule				
\multirow{5}*{12}	  	& $\V{\beta}_0$ & 0 	& -0.018 & 0.128 & 0.017 & 0.002 & 0.192 & 0.037 \\ 
  				& $\V{\beta}_1$ 	& 1 & 0.153 & 0.325 & 0.823 & 1.054 & 0.476 & 0.229 \\ 
  				& $\V{\beta}_2$ 	& 0.5 & -0.201 & 0.336 & 0.604 & 0.533 & 0.460 & 0.212 \\ 
  				& $\V{\beta}_3$ 	& 1 & 0.158 & 0.316 & 0.808 & 1.069 & 0.478 & 0.233 \\  
  				& $\V{\beta}_4$ 	& 2 & 0.906 & 0.317 & 1.297 & 2.163 & 0.567 & 0.347 \\ 		
\midrule	
\multirow{5}*{24}	& $\V{\beta}_0$ & 0 	& -0.011 & 0.124 & 0.016 & 0.002 & 0.192 & 0.037 \\ 
  				& $\V{\beta}_1$ 	& 1 & -0.088 & 0.330 & 1.293 & 1.054 & 0.476 & 0.229 \\ 
  				& $\V{\beta}_2$ 	& 0.5 & -0.431 & 0.331 & 0.975 & 0.533 & 0.460 & 0.212 \\ 
  				& $\V{\beta}_3$ 	& 1 & -0.086 & 0.315 & 1.279 & 1.069 & 0.478 & 0.233 \\ 
  				& $\V{\beta}_4$ 	&  2 & 0.641 & 0.324 & 1.952 & 2.163 & 0.567 & 0.347 \\  \bottomrule
\end{tabular}
\end{table}

\newpage

%The true parameter value is $\V{\theta} = (0, 1, 0.5,1,2)\Tra$.
\begin{table}[h!]
\caption[Effect of varying the number of outliers at a fixed location]{
Effect of varying the number of outliers at a fixed location.   
%The L$_2$E estimates  are essentially
%          unbiased for any number of outliers, while the MLE results become increasingly biased as outliers
%          are added.
\label{tab:nc}}
\centering
\begin{tabular}{ccccccccc}\\
\toprule
				&		&	& \multicolumn{3}{c}{MLE} & \multicolumn{3}{c}{L$_2$E} \\
Number of Outliers	& Coefficient & True Value &	mean 	& std & mse & mean & std & mse\\
\midrule
\multirow{5}*{0}		& $\V{\beta}_0$ & 0 		& 0.005 & 0.182 & 0.033 & 0.002 & 0.192 & 0.037 \\ 
						& $\V{\beta}_1$ 	& 1 & 1.026 & 0.433 & 0.188 & 1.054 & 0.476 & 0.229 \\ 
						& $\V{\beta}_2$ 	& 0.5 & 0.521 & 0.422 & 0.179 & 0.533 & 0.460 & 0.212 \\ 
						& $\V{\beta}_3$ 	& 1 & 1.041 & 0.438 & 0.193 & 1.069 & 0.478 & 0.233 \\ 
						& $\V{\beta}_4$ 	& 2 & 2.099 & 0.485 & 0.245 & 2.163 & 0.567 & 0.347 \\ 		 
\midrule				
\multirow{5}*{1}		& $\V{\beta}_0$  & 0 	& -0.022 & 0.157 & 0.025 & 0.002 & 0.192 & 0.037 \\ 
						& $\V{\beta}_1$ 	& 1 & 0.732 & 0.368 & 0.207 & 1.054 & 0.476 & 0.229 \\ 
						& $\V{\beta}_2$ 	& 0.5 & 0.296 & 0.369 & 0.178 & 0.533 & 0.460 & 0.212 \\ 
						& $\V{\beta}_3$ 	& 1 & 0.743 & 0.368 & 0.201 & 1.069 & 0.478 & 0.233 \\ 
						& $\V{\beta}_4$ 	& 2 & 1.662 & 0.392 & 0.268 & 2.163 & 0.567 & 0.347 \\ 
\midrule		
\multirow{5}*{5}		& $\V{\beta}_0$ & 0 	& -0.090 & 0.126 & 0.024 & 0.002 & 0.192 & 0.037 \\ 
						& $\V{\beta}_1$	& 1	 & 0.086 & 0.320 & 0.937 & 1.054 & 0.476 & 0.229 \\ 
						& $\V{\beta}_2$	& 0.5	& -0.263 & 0.327 & 0.689 & 0.533 & 0.460 & 0.212 \\ 
						& $\V{\beta}_3$	& 1 	& 0.090 & 0.308 & 0.922 & 1.069 & 0.478 & 0.233 \\ 
						& $\V{\beta}_4$	& 2	 & 0.830 & 0.312 & 1.466 & 2.163 & 0.567 & 0.347 \\ 						
\midrule				
\multirow{5}*{10}	& $\V{\beta}_0$ 	& 0 & -0.110 & 0.124 & 0.027 & 0.002 & 0.192 & 0.037 \\  
						& $\V{\beta}_1$	& 1	& -0.073 & 0.330 & 1.261 & 1.054 & 0.476 & 0.229 \\ 
						& $\V{\beta}_2$	& 0.5	& -0.417 & 0.333 & 0.951 & 0.533 & 0.460 & 0.212 \\ 
						& $\V{\beta}_3$	& 1	& -0.071 & 0.315 & 1.246 & 1.069 & 0.478 & 0.233 \\ 
						& $\V{\beta}_4$	& 2	& 0.659 & 0.323 & 1.903 & 2.163 & 0.567 & 0.347 \\ 
\midrule				
\multirow{5}*{15}	& $\V{\beta}_0$ 	& 0 & -0.117 & 0.124 & 0.029 & 0.002 & 0.192 & 0.037 \\ 
						& $\V{\beta}_1$	& 1	& -0.127 & 0.335 & 1.382 & 1.054 & 0.476 & 0.229 \\ 
						& $\V{\beta}_2$	& 0.5	& -0.470 & 0.338 & 1.055 & 0.533 & 0.460 & 0.212 \\ 
						& $\V{\beta}_3$	& 1	& -0.125 & 0.321 & 1.367 & 1.069 & 0.478 & 0.233 \\ 
						& $\V{\beta}_4$	& 2	& 0.605 & 0.328 & 2.054 & 2.163 & 0.567 & 0.347 \\ 				
\midrule				
\multirow{5}*{20}	& $\V{\beta}_0$ 	 & 0 & -0.122 & 0.124 & 0.030 & 0.002 & 0.192 & 0.037 \\ 
		  				& $\V{\beta}_1$	& 1	& -0.159 & 0.339 & 1.457 & 1.054 & 0.476 & 0.229 \\
		  				& $\V{\beta}_2$	& 0.5& -0.502 & 0.342 & 1.120 & 0.533 & 0.460 & 0.212 \\ 
		  				& $\V{\beta}_3$	& 1	& -0.157 & 0.325 & 1.443 & 1.069 & 0.478 & 0.233 \\
		  				& $\V{\beta}_4$	& 2	& 0.573 & 0.332 & 2.145 & 2.163 & 0.567 & 0.347 \\  \bottomrule
\end{tabular}
\end{table}

\section{Variable Selection Experiments in High Dimensions}

\begin{figure}
	\centering
	\includegraphics[scale=0.75]{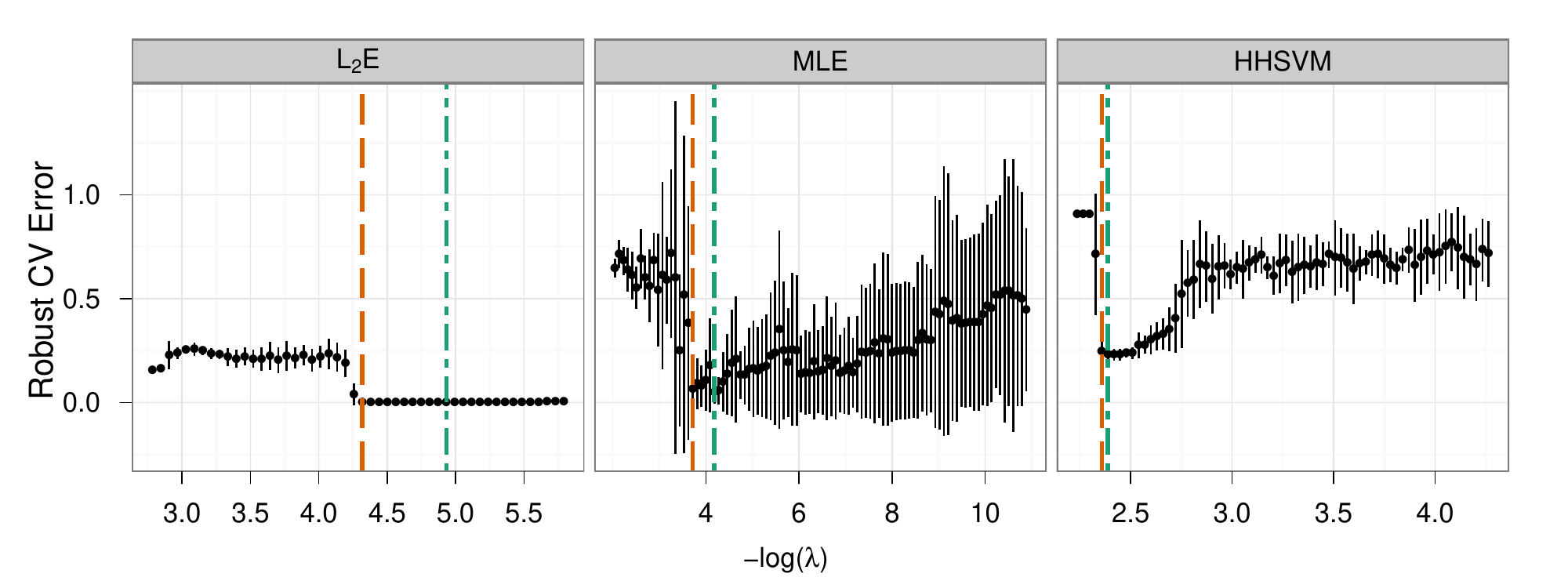}
	\caption{Robust 10-fold cross-validation curves for the three methods. The vertical error bars around the dots indicate $\pm$ one median absolute deviation with a scale factor of $1.4826$. The dash-dotted line indicates the minimizing $\lambda$. The dashed line indicates the 1-MAD rule $\lambda$.\label{fig:sim_cv_curves}}
\end{figure}

\begin{figure}
	\centering
	\includegraphics[scale=0.75]{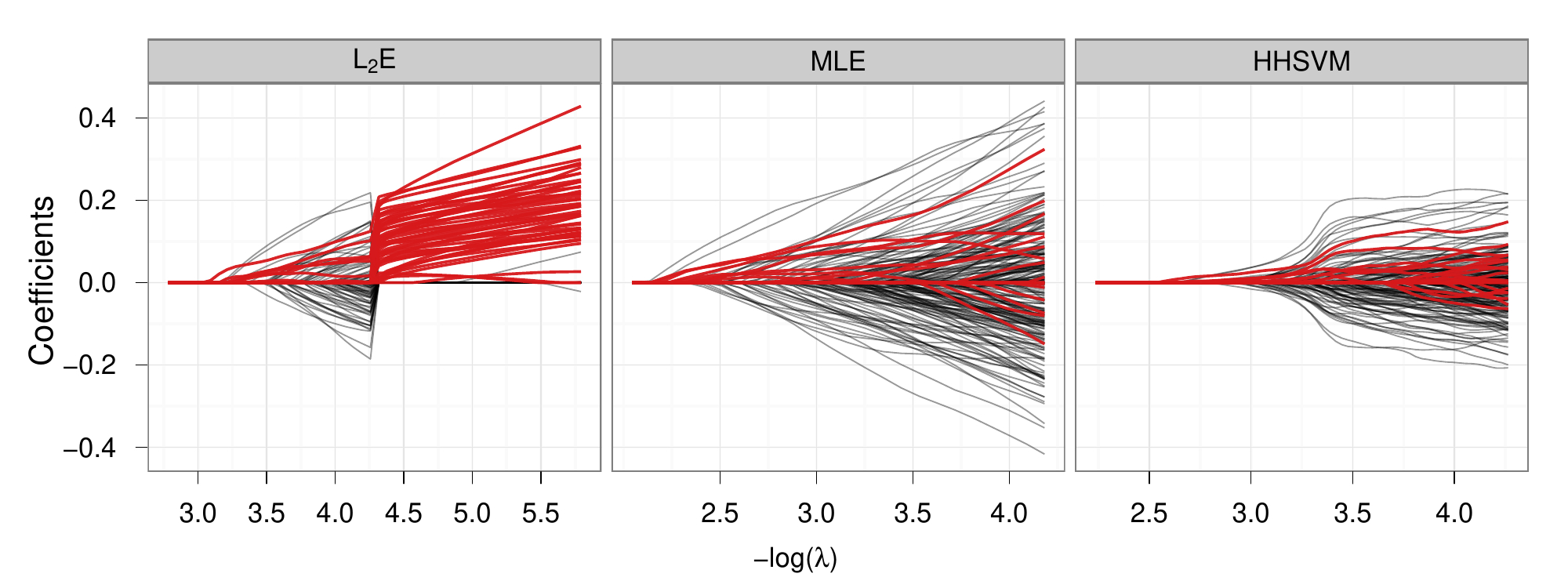}
	\caption{Regularization paths for the three methods. Paths for nonzero regression coefficients in the true model are drawn in heavy solid lines.\label{fig:sim_regularization_paths}}
\end{figure}

We show more detailed results for a single replicate for the simulations reported in Section 5.2.
Figure~\ref{fig:sim_cv_curves} shows the robust cross validation curves for the three methods for the replicate.
Figure~\ref{fig:sim_regularization_paths} shows the regularization paths for the three methods for the replicate.
Note the large jump in the L$_2$E curve. By choosing the starting L$_2$E point by our heuristic, a local minimum different from the MLE solution is found. For sufficiently large $\lambda$, however, the local minimum vanishes, and the regularization paths mimic the MLE regularization paths.

\newpage

\section{The Hybrid Huberized SVM}

Consider the following classification problem. Let $\M{X} \in \Real^{n \times p}$ denote a centered matrix of covariates and $\V{y} \in \{-1,1\}^n$ denote
binary class labels. We will employ the compact notation  $\Mtilde{X} = (\V{1}, \M{X}) \in \Real^{n \times (p+1)}$ and 
$\V{\theta} = (\beta_0, \V{\beta}\Tra)\Tra \in \Real^{p+1}$. The Hybrid Huberized Support Vector Machine (HHSVM) \citep{Wang2008} constructs a linear classifier $\Mtilde{X}\V{\theta}$ by minimizing the following loss.

\begin{equation*}
	\ell(\V{y}, \M{X}; \V{\theta}) =  \sum_{i=1}^n \phi\left (\VE{Y}{i} \Vtilde{x}_i\Tra\V{\theta} \right ) + J(\V{\beta}),
\end{equation*}
where the function $\phi$ is a smooth hinge loss,
\begin{equation*}
	\phi(u) = \begin{cases}
		(1 - t)^2 + 2(1-t)(t - u), 	& \text{if $u \leq t$,} \\
		(1 - u)^2,				& \text{if $t < u \leq 1$,} \\
		0,					& \text{otherwise,} \\
	\end{cases}
\end{equation*}
and $J$ is the Elastic Net penalty \citep{Zou2005}.
\begin{equation*}
J(\V{\beta}) = \lambda \left (\alpha \lVert \V{\beta} \rVert_1 + \frac{1-\alpha}{2} \lVert \V{\beta} \rVert_2^2 \right ),
\end{equation*}
where $\alpha \in [0, 1]$ is a mixing parameter between the 1-norm and 2-norm regularizers. We now derive an MM algorithm for solving the entire regularization path with respect to a varying $\lambda$ for a fixed $\alpha$.
The majorization we will use leads to a simple MM algorithm. This algorithm calculates a different regularization path than the algorithm in \citep{Wang2008}, which uses the following parameterization of the Elastic Net
\begin{equation*}
J(\V{\beta}) = \lambda_1 \lVert \V{\beta} \rVert_1 + \frac{\lambda_2}{2} \lVert \V{\beta} \rVert_2^2,
\end{equation*}
for  varying $\lambda_1$ for a fixed $\lambda_2$.  The code used in \citep{Wang2008} is available on the author's website (\url{http://www.stat.lsa.umich.edu/~jizhu/code/hhsvm}).

\newpage

\subsection{An MM Algorithm for Minimizing the Smooth Hinge Loss}

We begin by deriving a quadratic majorization of $\phi$. It is straightforward to verify that the first and second derivatives of $\phi$ are given by

\begin{equation*}
\begin{split}
	\phi'(u) &= \begin{cases}
		-2(1-t), 				& \text{if $u \leq t$,} \\
		-2(1 - u),				& \text{if $t < u \leq 1$,} \\
		0,					& \text{otherwise.} \\
	\end{cases} \\
	\phi''(u) &= \begin{cases}
		0,	 				& \text{if $u \leq t$,} \\
		2,					& \text{if $t < u \leq 1$,} \\
		0,					& \text{otherwise.} \\
	\end{cases}	 \\
\end{split}
\end{equation*}
Then we can express $\phi$ as an exact second order Taylor expansion at a point $\tilde{u}$ with
\begin{equation*}
	\phi(u) = \phi(\tilde{u}) + \phi'(\tilde{u})(u - \tilde{u}) + \frac{1}{2}\phi''(u^*)(u - \tilde{u})^2,
\end{equation*}
where $u^* = \delta u + (1-\delta) \tilde{u}$ for some $\delta \in (0,1)$. It follows immediately that
the following function majorizes $\phi$ at $\tilde{u}$.
\begin{equation*}
	g(u; \tilde{u}) = \phi(\tilde{u}) + \phi'(\tilde{u})(u-\tilde{u}) + (u-\tilde{u})^2.
\end{equation*}

The $u$ that minimizes $g(u; \tilde{u})$ is 
\begin{equation*}
\begin{split}
	u &= \tilde{u} - \frac{1}{2}\phi'(\tilde{u}) \\
	&= \tilde{u} + \left [
		(1 - t)I(u \leq t) + (1 - u)I(u > t)I(u \leq 1)
	\right ] \\
	&= \tilde{u} + 1 - \min (\max (\tilde{u}, t), 1) \\
\end{split}
\end{equation*}

%Figure~\ref{fig:MM_HHSVM} shows a sequence of majorizations corresponding to six MM iterates.
%
%\begin{figure}
%\centering
%\includegraphics[scale=0.65]{MM_HHSVM_Plots.pdf}
%\caption{A sequence of majorizations corresponding to six MM iterates. Notice that the curvature of the quadratic majorizations matches the curvature
%of the smooth hinge loss elbow.}
%\label{fig:MM_HHSVM}
%\end{figure}

\subsection{An MM Algorithm for the Unregularized Classification Problem}
Returning to our original problem and applying the above results along with the chain rule gives us the relationship 

\begin{equation*}
	\ell(\V{y}, \Mtilde{X}; \V{\theta}) \leq \ell(\V{y}, \Mtilde{X}; \Vtilde{\theta}) + \Vtilde{\varphi}\Tra\Mtilde{X}(\V{\theta} - \Vtilde{\theta}) + \lVert \Mtilde{X}(\V{\theta} - \Vtilde{\theta} )\rVert_2^2,
\end{equation*}
where
\begin{equation*}
	\Vtilde{\varphi}_i = \VE{y}{i}\varphi'(\VE{Y}{i}\Vtilde{x}_i\Tra\Vtilde{\theta}).
\end{equation*}
Since the equality occurs when $\V{\theta} = \Vtilde{\theta}$, the right hand side majorizes the left hand side. Furthermore, the majorization up to an additive constant is separable in $\beta_0$ and $\V{\beta}$.
\begin{equation*}
\begin{split}
	\left \lVert \frac{1}{2} \Vtilde{\varphi} + \Mtilde{X}(\V{\theta} - \Vtilde{\theta}) \right \rVert_2^2 &= 
	\left \lVert (\Mtilde{X}\Vtilde{\theta} - \frac{1}{2} \Vtilde{\varphi}) - \Mtilde{X}\V{\theta} \right \rVert_2^2  \\
	&= 	\left \lVert \left [\M{X}\Vtilde{\beta} - \frac{1}{2} (\Vtilde{\varphi} - \overline{\varphi}\V{1} ) 
		- \M{X}\V{\beta} \right ]
	+ \left [\tilde{\beta}_0\V{1} - \frac{1}{2}\overline{\varphi}\V{1} 
	- \beta_0\V{1} \right ] \right \rVert_2^2  \\
	&= 	
	n \left (\tilde{\beta}_0 - \beta_0 - \frac{1}{2n}\V{1}\Tra\Vtilde{\varphi} \right )^2 + 
	\left \lVert \Vtilde{z} - \M{X}\V{\beta} \right \rVert_2^2,
\end{split}
\end{equation*}
where
\begin{equation*}
	\Vtilde{z} = \M{X}\Vtilde{\beta} - \frac{1}{2} \left (\Vtilde{\varphi} - \frac{1}{n}\V{1}\Tra\Vtilde{\varphi} \V{1} \right).
\end{equation*}

%If $\Mtilde{X}$ is full rank then the update is
%
%\begin{equation*}
%	\V{\theta} = \Vtilde{\theta} + \frac{1}{2} \left (\Mtilde{X}\Tra\Mtilde{X} \right )^{-1} \Mtilde{X}\Tra \Vtilde{\varphi}.
%\end{equation*}

We can write the updates with the intercept and regression coefficients separately. The intercept update is
\begin{equation*}
	\beta_0 = \tilde{\beta}_0 - \frac{1}{2n}\V{1}\Tra\Vtilde{\varphi}.
\end{equation*}

and if $\M{X}$ is full rank the update for $\V{\beta}$ is

\begin{equation*}
	\V{\beta} = \Vtilde{\beta} - \frac{1}{2} \left (\M{X}\Tra\M{X} \right )^{-1} \M{X}\Tra \left (\Vtilde{\varphi} - \frac{1}{n}\V{1}\Tra\Vtilde{\varphi}\V{1} \right).
\end{equation*}

\subsection{An MM Algorithm for the HHSVM}

Adding an Elastic Net penalty to the majorization gives us the following loss function to minimize.

\begin{equation*}
	\frac{1}{2} \left (\tilde{\beta}_0 - \beta_0 - \frac{1}{2n}\V{1}\Tra\Vtilde{\varphi} \right )^2 + 
	\frac{1}{2n} \left \lVert \Vtilde{z} - \M{X}\V{\beta} \right \rVert_2^2 +
	\lambda \left (\alpha \lVert \V{\beta} \rVert_1 + \frac{1-\alpha}{2} \lVert \V{\beta} \rVert_2^2 \right ).
\end{equation*}

Penalized least squares problems of this variety are efficiently solved with coordinate descent. The coordinate descent updates are
\begin{equation*}
	\beta_j = \frac{S \left (\frac{1}{n}\V{x}_k\Tra \V{r}, \lambda\alpha \right )}
	{\frac{1}{n} \lVert \V{X}_k \rVert_2^2 + \lambda (1 - \alpha)},
\end{equation*}
where

\begin{equation*}
	\V{r} = \Mtilde{X}\Vtilde{\theta} - \frac{1}{2} \Vtilde{\varphi}  - \sum_{j \not = k} \VE{\beta}{j}\V{x}_j.
\end{equation*}

\bibliographystyle{asa}
\bibliography{references}

\end{document}